\input harvmac
\input epsf
%
%

\def\eg{{\it e.g.\/}}
\def\ie{{\it i.e.\/}}

\def\etal{{\it et al.\/}}

\def\frac#1#2{{{#1}\over {#2}}}

\def\smallfrac#1#2{\hbox{${{#1}\over {#2}}$}}

\def\MS{\hbox{$\overline{\rm MS}$}}
\def\AB{\hbox{$\rm AB$}}

\catcode`@=11 
\def\slash#1{\mathord{\mathpalette\c@ncel#1}}
 \def\c@ncel#1#2{\ooalign{$\hfil#1\mkern1mu/\hfil$\crcr$#1#2$}}
\def\lsim{\mathrel{\mathpalette\@versim<}}
\def\gsim{\mathrel{\mathpalette\@versim>}}
 \def\@versim#1#2{\lower0.2ex\vbox{\baselineskip\z@skip\lineskip\z@skip
       \lineskiplimit\z@\ialign{$\m@th#1\hfil##$\crcr#2\crcr\sim\crcr}}}
\catcode`@=12 

\def\PR{{\it Phys.~Rev.~}}
\def\PRL{{\it Phys.~Rev.~Lett.~}}
\def\NP{{\it Nucl.~Phys.~}}
\def\NPBPS{{\it Nucl.~Phys.~B (Proc.~Suppl.)~}}
\def\PL{{\it Phys.~Lett.~}}
\def\PRep{{\it Phys.~Rep.~}}

\def\ZP{{\it Zeit.~Phys.~}}

\def\vol#1{{\bf #1}}\def\vyp#1#2#3{\vol{#1} (#2) #3}
\def\as{\alpha_s}
%
%
\noblackbox
\pageno=0\nopagenumbers\tolerance=10000\hfuzz=5pt
\baselineskip 12pt
\line{\hfill {\tt hep-ph/0101192}}
\line{\hfill CERN-TH/2000-377}
\line{\hfill GeF/TH/8-00}
\line{\hfill RM3-TH/00-20}
\vskip 12pt
\centerline{\titlefont  Polarized Parton Distributions}
\vskip 12pt
\centerline{\titlefont from Charged--Current Deep-Inelastic Scattering}
\vskip 10pt
\centerline{\titlefont and Future Neutrino Factories}
\vskip 18pt
\centerline {Stefano Forte\footnote{$^\dagger$}{On leave from INFN,
Sezione di Torino, Italy}}
\vskip 6pt
\centerline {\it INFN, Sezione di Roma III}
\centerline {\it Via della Vasca Navale 84, I-00146 Rome, Italy}
\vskip 12pt
\centerline{Michelangelo L.~Mangano and Giovanni
Ridolfi\footnote{$^*$}{On leave from INFN,
Sezione di Genova, Italy}}
\vskip 6pt
\centerline{\it Theory Division, CERN}
\centerline{\it CH-1211 Geneva 23, Switzerland}
\vskip 52pt
\centerline{\bf Abstract}
{\narrower\baselineskip 10pt
\medskip\noindent
We discuss the determination of polarized parton distributions from
charged--current deep--inelastic scattering experiments.  We
summarize the next-to-leading order treatment of charged--current
polarized structure functions, their relation to polarized parton
distributions and scale dependence, and discuss their description by
means of a next-to-leading order evolution code. We discuss current
theoretical expectations and positivity constraints on the unmeasured
C--odd combinations $\Delta q-\Delta \bar q$ of polarized quark
distributions, and their determination in charged--current
deep--inelastic scattering experiments.  We give estimates of the
expected errors on charged--current structure functions at a future
neutrino factory, and perform a study of the accuracy in the
determination of polarized parton distributions that would be possible
at such a facility.
We show that these measurements have the potential to distinguish
between different theoretical scenarios for the proton spin structure.
\smallskip}
\vfill
\line{CERN-TH/2000-377\hfill}
\line{January 2001\hfill}
\eject \footline={\hss\tenrm\folio\hss}
%
%
\lref\topsup{G.~M.~Shore and G.~Veneziano, \PL \vyp{B244}{1990}{75}; 
\NP \vyp{B381}{1992}{23}; see
also G.~M.~Shore, {\tt hep-ph/9812355}.}
\lref\sutr{N.~W.~Park, J.~Schechter and H.~Weigel,
\PL\vyp{B228}{1989}{420}\semi
N.~W.~Park, J.~Schechter and H.~Weigel, \PR\vyp{D41}{1990}{2836}\semi
B.~Ehrnsperger and A.~Sch\"afer,
\PL\vyp{B348}{1995}{619}\semi
J.~Lichtenstadt and H.~J.~Lipkin,
\PL\vyp{B353}{1995}{119}\semi
J.~Dai \etal, {\it Phys. Rev.} {\bf D53} (1996) 273\semi
P.~G.~Ratcliffe, {\it Phys. Lett.} {\bf B365} (1996) 383.}
\lref\newdata{SMC Coll., B.~Adeva \etal, \PR\vyp{D58}{1998}{112001}  \semi
E155 Coll. P.~L.~Anthony \etal, \PL\vyp{B493}{2000}{19}.}
\lref\leader{M.~Anselmino, A.~Efremov and E.~Leader, \PRep\vyp{261}{1995}{1}.}
\lref\semexp{SMC Coll., B.~Adeva \etal, \PL\vyp{B420}{1998}{180}\semi
HERMES Coll, K.~Ackerstaff \etal, \PL\vyp{B464}{1999}{123}.}
\lref\semifit{D.~de~Florian and R.~Sassot,
\PR\vyp{D62}{2000}{094025}.}
\lref\glumod{M.~Gl\"uck \etal, {\tt hep-ph/0011215}.}
\lref\pos{G.~Altarelli, S.~Forte and G.~Ridolfi,
\NP\vyp{B534}{1998}{277}; \NPBPS\vyp{74}{1999}{138}.}
\lref\nnlovvn{W.~L.~van~Neerven and A.~Vogt, \NP\vyp{B568}{2000}{263}.}
\lref\anom{G.~Altarelli and G.~G.~Ross, \PL\vyp{B212}{1988}{391}\semi
R.~D.~Carlitz,
J.~C.~Collins and A.~H.~Mueller,\PL\vyp{B214}{1988}{229}\semi
G.~Altarelli and B.~Lampe, \ZP\vyp{C47}{1990}{315}\semi
W.~Vogelsang, \ZP\vyp{C50}{1991}{275}.}
\lref\koda{J.~Kodaira \etal, \PR\vyp{D20}{1979}{627};
\NP\vyp{B159}{1979}{99}\semi J.~Kodaira, \NP\vyp{B165}{1979}{129}.}
\lref\ap{G.~Altarelli and G.~Parisi, \NP\vyp{B126}{1977}{298}.}
\lref\dicder{D.~A.~Dicus, \PR\vyp{D5}{1972}{1367}\semi E.~Derman,
\PR\vyp{D7}{1973}{2755}.}
\lref\cgr{C.~G.~Callan and D.~J.~Gross, \PRL\vyp{22}{1969}{156}.}
\lref\ballrev{R.~D.~Ball, D.~A.~Harris and K.~S.~McFarland, {\tt
hep-ph/0009223}. }
\lref\spinrevth{See \eg\ 
G.~Ridolfi, {\it Nucl. Phys.} {\bf A666} (2000) 278\semi
R.~D.~Ball and
H.~A.~M.~Tallini, {\tt hep-ph/9812383}\semi
S.~Forte, {\tt hep-ph/9409416}, {\tt hep-ph/9610238}.}
\lref\spinrevexp{See \eg\ 
E.~W.~Hughes and R.~Voss, {\it  Ann. Rev. Nucl. Part. Sci.} {\bf 49}
(1999)  303\semi R.~Windmolders, {\tt hep-ph/9905505}.}
\lref\nufactrev{D.~M.~Kaplan, {\tt physics/0001037}\semi B.~J.~King,
{\tt hep-ex/9911008}, {\tt hep-ex/0001043}}
\lref\mucoll{The Muon Collider Coll., {\it $\mu^+\mu^-$
Collider: a feasibility study}, BNL-52503, Fermilab-Conf-96/092,
LBNL-38946\semi
B.~Autin, A.~Blondel and J.~Ellis eds., {\it Prospective Study of
Muon Storage Rings at CERN}, CERN 99-02, ECFA 99-197.}
\lref\nufactzz{See for example the studies presented at the
NuFACT'00 Workshop, May 22-26, Monterey, CA, {\tt
http://www.lbl.gov/Conferences/nufact00/}.}
\lref\nudisrev{J.~M.~Conrad, M.~H.~Shaevitz and T.~Bolton, {\it
Rev. Mod. Phys.} {\bf 70} (1998) 1341.}
\lref\emc{EMC Collaboration, J.~Ashman \etal, \PL\vyp{B206}{1988}{364};
                \NP\vyp{B328}{1989}{1}.}
\lref\cracovia{G.~Altarelli, R.D.~Ball, S.~Forte and G.~Ridolfi,
               \NP\vyp{B496}{1997}{337}\semi
G.~Altarelli, R.~D.~Ball, S.~Forte and
G.~Ridolfi, {\it Acta Phys. Pol.} {\bf B29} (1998) 1145.}
\lref\smcfit{ SMC Collaboration, B. Adeva \etal, \PR\vyp{D58}{1998}{112002}.}
\lref\strange{S.~J.~Brodsky, J.~Ellis and M.~Karliner, \PL \vyp{B206}{1988}{309}\semi
J.~Ellis and  M.~Karliner, {\tt  hep-ph/9601280}.}
\lref\glue{S.~Forte, {\it Acta Phys. Pol.} \vyp{B22}{1991}{1065}
\semi
A.~H.~Mueller, \PL\vyp{B234}{1990}{517}.} 
\lref\cchera{See \eg\ T. Benisch, H1 and ZEUS,
\NP\vyp{A666}{2000}{141}.}
\lref\ccpolhera{M.~Anselmino, P.~Gambino and J.~Kalinowski,
\PR\vyp{D55}{1997}{5841}.}
\lref\agk{M.~Anselmino, P.~Gambino and J.~Kalinowski,
\ZP\vyp{C64}{1994}{267}.}
\lref\ravi{V.~Ravishankar, \NP\vyp{B374}{1992}{309}.}
\lref\ope {M.~Maul \etal, \ZP\vyp{A356}{1997}{443}.}
\lref\blum{J.~Bl\"umlein and N.~Kochelev, \NP\vyp{B498}{1997}{285}.}
\lref\deflo{D.~de~Florian and R.~Sassot, \PR\vyp{D51}{1995}{6052}.}
\lref\kretz{S.~Kretzer and M.~Stratmann, {\it Eur.~Phys.~Jour.}
\vyp{C10}{1999}{107}.}
\lref\nload{R.~Mertig and W.~L.~van~Neerven, \ZP\vyp{C70}{1996}{637}\semi
              W.~Vogelsang, \PR\vyp{D54}{1996}{2023}, \NP\vyp{B475}{1996}{47}.}
\lref\kis{R.~K.~Ellis, W.~J.~Stirling and B.~R.~Webber, {\it QCD and
Collider Physics} (Cambridge U.P., Cambridge 1996).}
\lref\intc{S.~J.~Brodsky \etal, \PL\vyp{B93}{1980}{451}.}
\lref\vn{E.~B.~Zijlstra and  W.~L.~van~Neerven,
\NP\vyp{B417}{1994}{61}}
\lref\bfr{R.D.~Ball, S.~Forte and G.~Ridolfi,
\PL\vyp{B378}{1996}{255}.}
\lref\guidorev{See \eg\ G.~Altarelli, {\it Phys. Rep.}
\vyp{81}{1982}{1}.}
\lref\rosac{D.~A.~Ross and C.~T.~Sachrajda, \NP\vyp{B179}{1979}{497}.}
\lref\nspol{M.~Stratmann, A.~Weber and W.~Vogelsang, \PR\vyp{D53}{1996}{238}.}
\lref\bpz{V.~Barone, C.~Pascaud and F.~Zomer, {\it Eur.~Phys.~Jour.} \vyp{C12}{2000}{243}.}
\lref\cteq{CTEQ Coll., H.~L.~Lai \etal, 
    {\it Eur.~Phys.~Jour.} \vyp{C12}{2000}{375}.}
\lref\inst{S.~Forte, \PL\vyp{B224}{1989}{189}; \NP\vyp{B331}{1990}{1}\semi
S.~Forte and E.~V.~Shuryak, \NP\vyp{B357}{1991}{153}.}
\lref\ints{S.~J.~Brodsky and B.-Q.~Ma, \PL\vyp{B381}{1996}{317}.}
\lref\Geer{S.~Geer,
{\it Phys. Rev.}  {\bf D57} (1998) 6989, Erratum {\it ibid.},
{\bf D59} (1999) 039903.}
\lref\bkbook{ I. Bigi \etal,
{\it The potential for neutrino physics at muon colliders 
and dedicated high current muon storage rings},  
BNL-67404.  }
\lref\Albright{
C.~Albright \etal,
{\it Physics at a neutrino factory},
{\tt hep-ex/0008064.}}
\lref\nufactexp{D.A. Harris, and K.S.McFarland, {\tt hep-ex/9804010}, 
proceedings of the 
Workshop on Physics at the First Muon Collider and
at the Front End of a Muon Collider, November 1997, Fermilab. }
\lref\Lai{CTEQ Coll., H.~L.~Lai \etal,
{\it Phys. Rev.} {\bf D55} (1997) 1280 .}
\lref\cernrep{M.~L.~Mangano \etal, QCD/DIS Working Group, to appear in the 
Report of the CERN/ECFA Neutrino Factory Study Group.}
\lref\smctarg{SMC Coll., D.~Adams \etal, {\it Nucl. Instrum. Meth.}
{\bf A437} (1999) 23.}


\newsec{Introduction}

Polarized deep-inelastic scattering (DIS) has attracted considerable
theoretical~\spinrevth\ and experimental~\spinrevexp\ interest since
the EMC experimental results in 1988~\emc\ showed that the proton spin
structure is subtler than naive parton expectations might suggest.  A
first generation of experiments with electron and muon beams, and
proton, deuterium and neutron (\ie\ $^3$He) targets has provided us
with information on individual polarized parton distributions which
give important clues on the relevant underlying theoretical
issues. Several crucial pieces of information, however, are still
missing. Specifically, current inclusive DIS experiments only measure
the C--even combination $\Delta q+\Delta \bar q$ of quark
distributions.  Moreover, they are only very weakly sensitive to the
size of the strange quark distribution. Therefore, a full flavor
separation is very difficult, and a separate measurement of the quark
and antiquark components is impossible. Furthermore, there are
indications~\refs{\cracovia,\smcfit} that the polarized gluon
distribution is large, but uncertainties are still sizable. Several of
these pieces of information are crucial not only in reconstructing the
full spin structure of the nucleon, but also (perhaps more
interestingly) in deciding among the various theoretical scenarios
which have been proposed to understand the somewhat puzzling picture
which the original EMC results and their subsequent refinements point
at. In particular, it is crucial to determine accurately the size of
the strange contribution~\strange\ and of the polarized gluon~\glue.

In the unpolarized case, the most precise available information on the
strange quark distribution and the quark--antiquark separation comes
from charged--current deep-inelastic scattering~\nudisrev, which, being
mediated by a charged current, gives access to the flavor structure of
the target. In the polarized case, the luminosity of existing neutrino
beams requires targets whose typical sizes are of the orders of
several meters, and therefore cannot be polarized. In principle,
charged--current events can be a significant part of the cross section
even with electron or muon beams if the energy is high enough. Indeed,
charged--current unpolarized cross sections and structure functions
have been measured at HERA~\cchera, and the corresponding polarized
asymmetries have been studied~\ccpolhera\ as a way of gaining
information on polarized parton distributions if the polarized option
were to be available at HERA.

It is several years now since muon storage rings have been proposed as
a powerful new way to reach high energies and luminosities in a
colliding-beam facility~\refs{\mucoll,\nufactrev}.  It has since been
recognized that exciting physics can also be obtained with the highly
focused neutrino beams arising from the decays of muons along straight
sections of the accumulator~\Geer. In addition to the obvious
applications for studies of neutrino properties (masses and mixings)
using far-away detectors (for a recent detailed study see
Ref.~\Albright), DIS experiments operating close-by downstream the
muon ring could provide significant contributions to several
topics~\refs{\bkbook,\cernrep}: from the study of the nucleon
structure to accurate measurement of the parameters of weak and strong
interactions, from the determination of CKM matrix elements to studies
of heavy quark decays, and more.  Reference~\bkbook\ pointed out the
potential for measurements of unparalleled precision of both
unpolarized and polarized neutrino structure functions, leading to an
accurate decomposition of the partonic content of the nucleon in terms
of individual (possibly spin-dependent) flavor densities.  A more
quantitative investigation of polarized DIS with neutrino beams is
therefore called for, both in order to assess the physics potential of
these experiments, and also to decide among other possible options,
such as polarized HERA.  A first study in this direction has recently
appeared in~\ballrev.

Several studies of polarized neutrino DIS have been presented in the
literature. In particular, the parton model formalism has been
presented in Ref.~\agk, while its relation to perturbative QCD has
been worked out through the operator--product expansion at leading
order in Refs.~\refs{\ravi\ope{--}\blum}. The next-to-leading order coefficient
functions were computed several years ago~\deflo, and have been
recently generalized to the case where heavy quark masses are
retained~\kretz; next-to-leading order polarized anomalous dimensions
were determined for the full set of operators relevant to both neutral--
and charged--current DIS recently~\nload. Even though all the
theoretical tools which are needed in order to perform a full
next-to-leading order analysis are thus essentially available in the
literature, no such study has been performed yet.

In this paper, we wish to give a first systematic next-to-leading
order treatment of charged--current polarized DIS. The purpose of this
work is to provide a set of common theoretical tools, as well as some
benchmark estimates for future work.  We will start (Sect.~2) by
recalling the structure of the polarized asymmetries in
charged--current DIS in terms of structure functions, review their
leading--order expressions in terms of parton distributions, and the
scale dependence of the parton distributions. We will then recall how
charged--current lepton and antilepton DIS on proton and neutron
targets allows a full flavor decomposition of the nucleon spin
content.  In Sect.~3, we will discuss the full next-to-leading order
formalism for structure functions and parton distributions (this
section can be skipped by those who are only interested in the
phenomenological analysis).  The results presented here have been
incorporated in an upgrade of the ABFR~\cracovia\ next-to-leading
order evolution code for polarized parton distributions, by including
the option of evolving charged--current structure functions and
fitting them to experimental data. We will next (Sect.~4) discuss the
constraints that the elementary requirement of positivity of
cross--sections imposes on the otherwise unknown C--odd combinations
of quark and antiquark distributions $\Delta q_i -\Delta\bar q_i$, and
how they limit the set of possible expected results for these
quantities. Then, we will examine the current theoretical expectations
for the strange and quark-antiquark content of the nucleon within
several theoretical scenarios which have been suggested in order to
understand current information on the nucleon spin.

We will turn to phenomenology in Sect~5, where we will discuss
the precision on the determination of structure functions which is
expected at a future muon storage ring facility: in particular, we
will review the expected characteristics of the neutrino beam, and
present detailed estimates of the expected percentage error on
structure function determinations as a function of the kinematic
variables $x$ and $Q^2$. We will then (Sect.~6) study how such a
determination of polarized structure functions would improve our
knowledge of parton distributions: we will generate pseudo--data
according to the theoretical scenarios of Sect.~4, distributed
according to to the error estimates of Sect.~5, and use the evolution
code discussed in Sect.~3 to perform next-to-leading order fits of
structure functions to these data. We will thus be able to assess the
error on individual parton distributions which would be obtained
thanks to this kind of experiment.

\newsec{Flavor content and scale dependence of polarized structure functions}

Polarized structure functions for charged--current DIS have been
computed in the parton model in Ref.~\agk, and at leading order in
perturbative QCD with operator methods in
Refs.~\refs{\ope,\blum}. Here for completeness we briefly review the
derivation of the leading order expression of these structure
functions from the QCD improved parton model, the evolution equations
which govern their scale dependence, and then review how it is
possible to use this information to disentangle the polarized flavor
and antiflavor content of the target.

\subsec{Structure functions and parton distributions}
We define structure functions in terms of the
hadronic  tensor $W^{\mu\nu}$ as follows:
\eqn\hadten{\eqalign{W_{\mu\nu}=&\left(-g_{\mu\nu}+ {q_\mu q_\nu\over  
q^2}\right)F_1(x,Q^2)+ {\hat p_\mu\hat p_\nu\over p\cdot q}F_2(x,Q^2)
+i\epsilon_{\mu\nu\alpha\beta}
{q^\alpha p^\beta\over 2 p\cdot q} F_3(x,Q^2)\cr&\quad
-i\epsilon_{\mu\nu\alpha\beta} {q^\alpha s^\beta\over  p\cdot q} g_1(x,Q^2)
-i\epsilon_{\mu\nu\alpha\beta}{q^\alpha \left(p\cdot q \, s^\beta-s\cdot
q \, p^\beta\right)
\over ( p\cdot q)^2} g_2(x,Q^2)\cr&\qquad
+{1\over p\cdot
q}\left[{1\over 2} \left(\hat p_\mu \hat s_\nu+ \hat p_\nu\hat
s_\mu\right)-{s\cdot q\over p\cdot q}\hat p_\mu\hat p_\nu\right]
g_3(x,Q^2)\cr&\qquad\qquad+
{s\cdot q\over p\cdot q} \left[ {\hat p_\mu\hat p_\nu\over p\cdot q}
g_4(x,Q^2)+\left(-g_{\mu\nu}+{q_\mu q_\nu\over q^2}\right)g_5(x,Q^2)\right],
\cr}}
where $q^\mu$, $p^\mu$ and $s^\mu$ are respectively 
the momentum of the incoming virtual gauge bosons, and the momentum
and spin of the incoming nucleon, and
\eqn\hatpq{\hat p_\mu\equiv p_\mu -{p\cdot q\over q^2}q_\mu;\qquad
\hat s_\mu\equiv s_\mu -{s\cdot q\over q^2}q_\mu.}
The proton spin vector is normalized as $s^2=-m^2$,
where $m$ is the hadron mass. The hadronic tensor is defined as~\kis
\eqn\wmunudef
{W_{\mu\nu}={1\over 4\pi} \int d^4z\,e^{iq\cdot z}\,
\langle p,s| \left[ J_\mu(z),J^\dagger_\nu(0)\right]|p,s\rangle.}
Notice that the definition with $\mu\leftrightarrow \nu$ is sometimes
(\eg\ in~\agk) adopted; this corresponds to changing the sign of the
antisymmetric part of the tensor.
The definitions of the unpolarized structure functions $F_i$ and of
the polarized parity-conserving structure functions $g_1$, $g_2$ are
essentially standard, while a variety of different conventions have
been adopted in the literature for the polarized parity-violating
structure functions $g_3$, $g_4$ and $g_5$. Here we adopt the same convention
as Ref.~\blum.

We now specialize to the case of charged--current scattering. 
The total cross section for charged--current DIS on unpolarized
targets is given by
\eqn\xsectot
{{d^2\sigma^{\lambda_\ell}(x,y,Q^2)\over dx dy}=
{G^2_F  \over 2\pi (1+Q^2/m_W^2)^2}
{Q^2\over xy}
\left[-\lambda_\ell\, y \left(1-\smallfrac{y}{2}\right) x F_3
     + (1-y- x^2 y^2 \smallfrac{m^2}{Q^2} ) F_2 + y^2 x F_1\right]}
where $\lambda_\ell=\pm 1$ is the
helicity of the incoming lepton ($-1$ for a neutrino or electron,
$+1$ for an antineutrino or positron).
The explicit expressions of the various structure functions
in eq.~\xsectot\ depend only on the charge of the exchanged $W$, and on the
target.
The polarized cross-section difference
\eqn\deltadef{\Delta\sigma\equiv
\sigma(\lambda_p=-1)-\sigma(\lambda_p=+1),} 
where $\lambda_p=\pm 1$ is the proton helicity, is given by
\eqn\xsecas{\eqalign{
{d^2\Delta\sigma^{\lambda_\ell}(x,y,Q^2)\over dx dy}=&
{G^2_F  \over \pi (1+Q^2/m_W^2)^2}{Q^2\over xy}
\Bigg\{
  \left[
     -\lambda_\ell\, y (2-y)  x g_1- (1-y) g_4- y^2 x g_5
  \right]\cr&\quad
+2xy{m^2\over Q^2}
  \left[\lambda_\ell x^2y^2 g_1 +\lambda_\ell  2 x^2 y g_2
       +\left(1-y-x^2y^2\smallfrac{m^2}{Q^2}\right) xg_3
  \right.\cr&
\left.\qquad\quad
   -x\left(1-\smallfrac{3}{2}y-x^2 y^2 \smallfrac{m^2}{Q^2}\right)
g_4-x^2y^2 g_5 \right]
\Bigg\}.\cr}}
The results in eqs.~\xsectot\ and~\xsecas\ agree with those given in
Ref.~\agk, while they are by a factor of two larger than the results
of Ref.~\blum. The definition eq.~\deltadef\ corresponds to the
difference of antiparallel minus parallel spins of the incoming
particles for an incoming lepton, but parallel minus antiparallel
spins for an incoming antilepton.

Equation~\xsecas\ shows that for longitudinal nucleon polarization the
contributions of the structure functions $g_2$ and $g_3$ to the
cross--section are suppressed by powers of
$\smallfrac{m^2}{Q^2}$. These structure functions give an unsuppressed
contribution to the cross--section for transverse polarization, but in
such case the polarized cross section difference itself vanishes as
$\smallfrac{m^2}{Q^2}\to 0$. Henceforth, we will systematically
neglect all power suppressed contributions; therefore, we will not
discuss these structure functions further. It should be pointed out,
however, that at the neutrino factory it will be possible in principle
to extract the power-suppressed contributions directly from the data,
by studying the $Q^2$ dependence.

Because the same tensor structures appear in the spin-dependent and
spin-independent parts of the hadronic tensor eq.~\hadten\ in the
$\smallfrac{m^2}{Q^2}\to 0$ limit, the cross section difference
eq.~\xsecas\ is obtained from the total cross-section eq.~\xsectot\
replacing \eqn\polunpol{F_1\to -g_5,\quad F_2\to-g_4,\quad F_3\to 2
g_1} and multiplying by a factor two due to the fact that the total
cross--section is an average over initial state polarizations.  In
particular, it is easy to derive a polarized analogue~\dicder\ of the
Callan-Gross relation~\cgr, which follows from the observation that
the quark-gluon coupling conserves helicity when all masses are
neglected, so the hadronic tensor eq.~\hadten, if computed at leading
order, must vanish when contracted with a longitudinal polarization
vector $\epsilon_\nu^L$. Since $\epsilon_\nu^L$ can be written as a
linear combination of $p$ and $q$, this condition implies $p_\mu p_\nu
W^{\mu\nu}=0$, which using the expression eq.~\hadten\ becomes
\eqn\cgder{p_\mu p_\nu W^{\mu\nu}={(p\cdot q)^2\over
q^2}\left[\left(F_1- {F_2\over 2x}\right)+{s\cdot q\over p\cdot
q}\left(g_5-{g_4\over 2x}\right)\right]=0.}
Therefore at leading order 
\eqn\cgrel{g_4(x,Q^2)= 2 x g_5(x,Q^2),} 
and, at leading twist, there are only two independent polarized structure
functions, namely $g_1$ (parity conserving) and $g_5$ (parity
violating), respectively analogous to the unpolarized structure
functions $F_1$ and $F_3$. We will see in Sect.~3.1 that also
beyond leading order there are only two independent structure functions,
despite the fact that eq.~\cgrel\ is violated.

The leading--order expression of $g_1$ and $g_5$ are straightforwardly
found by computing the hadronic tensor for  $W^\pm$ scattering off a 
free quark~\agk:
\eqn\gglo{\eqalign{g_1^{W^+}(x,Q^2)&=
\Delta \bar u(x,Q^2)+\Delta d(x,Q^2)+ \Delta \bar c(x,Q^2)+\Delta s(x,Q^2)\cr
g_1^{W^-}(x,Q^2)&=
\Delta  u(x,Q^2)+\Delta \bar d(x,Q^2)+ \Delta c(x,Q^2)+\Delta \bar  s(x,Q^2)\cr
g_5^{W^+}(x,Q^2)&=
\Delta \bar u(x,Q^2)-\Delta d(x,Q^2)+ \Delta \bar c(x,Q^2)-\Delta s(x,Q^2)\cr
g_5^{W^-}(x,Q^2)&=
-\Delta u(x,Q^2)+\Delta \bar d(x,Q^2)- \Delta c(x,Q^2)+\Delta \bar  s(x,Q^2).
\cr}}
Below charm threshold only the Cabibbo-suppressed part of the $\Delta
s$ contribution survives, because the Cabibbo-enhanced transition
would require production of a $c$ quark in the final state. A $\Delta
c$ contribution is in principle possible even below charm threshold in
case intrinsic charm~\intc\ is present, since it only requires
production of a strange quark. Similarly, the Cabibbo-suppressed part
of the down contribution vanishes.  Hence, below charm threshold the
leading--order expression of the structure functions is
\eqn\gglobct{\eqalign{g_1^{W^+}(x,Q^2)&=\Delta \bar u(x,Q^2)+
\cos^2\theta_c\Delta d(x,Q^2)+ \Delta \bar c(x,Q^2)+\sin^2\theta_c
\Delta s(x,Q^2)\cr
g_1^{W^-}(x,Q^2)&=\Delta  u(x,Q^2)+\cos^2\theta_c\Delta \bar d(x,Q^2)+ \Delta
c(x,Q^2)+\sin^2\theta_c \Delta \bar  s(x,Q^2)\cr
g_5^{W^+}(x,Q^2)&=\Delta \bar u(x,Q^2)-\cos^2\theta_c\Delta d(x,Q^2)+ \Delta
\bar c(x,Q^2)-\sin^2\theta_c \Delta s(x,Q^2)\cr
g_5^{W^-}(x,Q^2)&=-\Delta u(x,Q^2)+\cos^2\theta_c\Delta \bar d(x,Q^2)- \Delta
c(x,Q^2)+\sin^2\theta_c \Delta \bar  s(x,Q^2),\cr}}
where $\theta_c$ is the Cabibbo angle, and $\Delta c$ is the intrinsic
charm distribution. The remaining mixing angles in
the CKM matrix can be taken to vanish for all practical purposes, so a
contribution from $\Delta b$ will only be present above the top
threshold; it will then have the form which straightforwardly follows
from repeating the pattern of previous generations.  Equation~\gglo\
is to be contrasted with the familiar case of virtual photon
scattering, where
\eqn\gnclo{\eqalign{ g_1^{\gamma^*}(x,Q^2)
&={1\over 2}\sum_{i=1}^{n_f} e^2_i \left[\Delta
q_i(x,Q^2)+\Delta \bar q_i(x,Q^2)\right]\cr
g_5^{\gamma^*}(x,Q^2)&=0.\cr}}

\subsec{Evolution equations}
The scale dependence of parton distributions is determined by the
Altarelli-Parisi evolution equations~\refs{\ap,\guidorev}.  Define the
C--even and C--odd quark distributions
\eqn\cevod{\Delta q^\pm_i=\Delta q_i \pm \Delta \bar{q}_i,}
the singlet combination
\eqn\qsdef{
\Delta \Sigma^\pm=\sum_{i=1}^{n_f}\Delta q_i^\pm; 
\qquad\Delta\Sigma\equiv\Delta\Sigma^+}
and the nonsinglet combination
\eqn\qnsdef{
\Delta q^{NS\,\pm}_{ij}=
\left(\Delta q_i^\pm-\Delta q_j^\pm\right ),\quad i\not=j.}
The evolution equations have the form
\eqn\apeq{\eqalign{
&\frac{d}{dt}\Delta q^{NS\,\pm}_{ij}
= \frac{\as(t)}{2\pi}\Delta P_{NS}^\pm \otimes \Delta q^{NS\,\pm}_{ij},\cr
&\frac{d}{dt}\Delta \Sigma^-
= \frac{\as(t)}{2\pi} \Delta P_{S}^- \otimes \Delta \Sigma^-,\cr
&\frac{d}{dt}\ \left (\matrix{\Delta \Sigma \cr \Delta g}\right)
=\frac{\as(t)}{2\pi}
\left (\matrix{\Delta P^{+}_S & 2n_f\Delta P_{qg} \cr \Delta P_{gq}
& \Delta P_{gg}}\right) \otimes 
\left (\matrix{\Delta \Sigma \cr \Delta g}\right),}}
where $t = \log{(Q^2/\Lambda^2)}$, $\otimes$ denotes the usual
convolution with respect to $x$, and all splitting functions admit a
perturbative expansion of the form $\Delta P(x,\as)=\Delta
P^{(0)}+{\as\over 2\pi} \Delta P^{(1)}+\dots$.

Notice that the gluon distribution cannot mix with the C--odd
distributions because charge conjugation implies that
\eqn\ccqg{P_{qg}=P_{\bar qg};\qquad P_{gq}=P_{g\bar q}.}
Furthermore, because all flavors are equivalent when masses are
neglected, it is only necessary to distinguish flavor-conserving
(diagonal) splittings $P^D_{qq}\equiv P_{q_iq_i}$ and flavor-changing
(non-diagonal) splittings $P^{ND}_{qq}\equiv P_{q_iq_j}$ with
$i\not=j$.  Finally, also by charge conjugation,
\eqn\ccqq{
 P_{q\bar q}=P_{\bar q q},\quad
P_{\bar q\bar q}=P_{qq};\qquad \Delta P_{q\bar q}=\Delta P_{\bar q q};\quad
\Delta P_{\bar q\bar q}=\Delta P_{qq}.} 
It then follows immediately that, in general
\eqn\pexp{\eqalign{\Delta P^{\pm}_{NS}&=\left(\Delta P^D_{qq}-\Delta P^{ND}_{qq}\right)\pm
\left(\Delta P^D_{q\bar q}-\Delta P^{ND}_{q\bar q}\right) \cr
\Delta P^{\pm}_{S}&=\left(\Delta P^D_{qq}+(n_f-1)\Delta P^{ND}_{qq}\right)\pm
\left(\Delta P^D_{q\bar q}+(n_f-1)\Delta P^{ND}_{q\bar q}\right),\cr}}
similarly to the unpolarized case (with $P\to\Delta P$).

Specializing now to the leading--order case, note that, because the
quark--gluon interaction conserves helicity, flavor, and baryon
number, at leading order there is only one quark-quark splitting
function
\eqn\lopqq{P^{\rm LO}_{q_iq_i}=\Delta P^{\rm LO}_{q_iq_i}\equiv
P^{\rm LO}_{qq},}
while 
\eqn\vanlop{\eqalign{P^{\rm LO}_{q\bar q}=&
\Delta P^{\rm LO}_{q\bar q}=0\cr
P^{ND,\,{\rm LO}}_{q q}=&\Delta 
P^{ND,\,{\rm LO}}_{q q}=0.}}
It follows that at leading order all quark 
(polarized and unpolarized) 
distributions evolve with the same splitting
function:
\eqn\pqqeq{\Delta P^{-}\equiv\Delta P^{-,\,{\rm LO}}_{NS}=\Delta P^{+,\,{\rm LO}}_{NS}=
\Delta P^{\pm,\,{\rm LO}}_{S}=P^{\rm LO}_{qq}.}

\subsec{Flavor decomposition}
Assuming the availability of neutrino and antineutrino beams, and the
capability of measuring independently $g_1$ and $g_5$, we have four
different linear combinations of individual polarized parton
distributions from charged--current scattering, on top of the usual
one (or two, if proton and neutron targets are available) from
neutral--current scattering. Using the leading order expressions of
the structure functions, eq.~\gglo, we get
\eqnn\flavcoma\eqnn\flavcomb 
$$\eqalignno{\smallfrac{1}{2}
\left(g_1^{W^-}-g_5^{W^-}\right)&=\Delta u+\Delta c;\qquad
\smallfrac{1}{2}
\left(g_1^{W^+}+g_5^{W^+}\right)=\Delta \bar u+\Delta \bar c;
&\flavcoma
\cr
\smallfrac{1}{2}\left(g_1^{W^+}-g_5^{W^+}\right)&=\Delta d+\Delta s;
\qquad
\smallfrac{1}{2}\left(g_1^{W^-}+g_5^{W^-}\right)=\Delta \bar
d+\Delta \bar s.
&\flavcomb\cr
}$$
Below charm threshold, in the absence of intrinsic charm, and
neglecting Cabibbo-suppressed contributions, the $\Delta c$ and
$\Delta s$ contributions decouple, and all four light flavors and
antiflavors can be extracted directly from eqs.~\flavcoma,\flavcomb.
Above charm threshold, or even below threshold if intrinsic charm is
present, a direct measurement of all individual parton distributions
is only possible if further independent linear combinations are
experimentally available.

For a nucleon target, further independent linear combinations of
parton distributions can be determined by using both proton and
neutron (or deuterium) targets: by isospin, all structure functions
for a neutron target are expressed in terms of parton distributions of
the proton by interchanging $u$ and $d$ in the expressions for a
proton target. This then gives us four new combinations, obtained
interchanging $u$ and $d$ in eqs.~\flavcoma,\flavcomb. Combining these
with the combinations eqs.~\flavcoma,\flavcomb, however, only gives us
six independent linear combinations. A convenient choice is for
instance
\eqnn\flavcomsnsa\eqnn\flavcomsnsb\eqnn\flavcomsnsc
$$\eqalignno{&\smallfrac{1}{2}
\left(g_1^{W^+}-g_5^{W^+}\right)[n-p]=\Delta u-\Delta d;\quad
\smallfrac{1}{2}
\left(g_1^{W^-}+g_5^{W^-}\right)[n-p]=\Delta \bar u-\Delta \bar d;
&\flavcomsnsa\cr
&\smallfrac{1}{2}\left(g_1^{W^-}-g_5^{W^-}\right)[p]-
\smallfrac{1}{2}\left(g_1^{W^+}-g_5^{W^+}\right)[n]=\Delta s-\Delta
c;\cr&\qquad\qquad\qquad\qquad
\smallfrac{1}{2}\left(g_1^{W^-}+g_5^{W^-}\right)[p]-\smallfrac{1}{2}
\left(g_1^{W^+}+g_5^{W^+}\right)[n]=\Delta \bar s-\Delta
\bar c;
&\flavcomsnsb\cr
&\smallfrac{1}{2}
\left(g_1^{W^+}-g_5^{W^+}\right)[p+n]=\Delta u+\Delta d+2\Delta s;\cr
&\qquad\qquad\qquad\qquad
\smallfrac{1}{2}
\left(g_1^{W^-}+g_5^{W^-}\right)[p+n]=\Delta \bar u+\Delta \bar
d+2\Delta \bar s.
&\flavcomsnsc\cr
}$$
The structure functions measured from virtual photon scattering
eq.~\gnclo\ do not provide any further independent linear combination
of parton distributions.

It follows that a complete separation of the four active flavors and
antiflavors at a fixed scale above charm threshold from inclusive
nucleon structure functions only is not possible. However, the
separation is possible by comparing structure functions above and
below threshold. For instance, neglecting Cabibbo-suppressed terms,
below charm threshold eq.~\flavcomb\ determines the down polarized
(quark and antiquark) distributions, eq.~\flavcomsnsb\ gives the
(intrinsic) charm, and then either eq.~\flavcoma\ or eq.~\flavcomsnsa\
give the up distribution, the other equation providing a consistency
check. Hence, the up, down and (intrinsic) charm can always be
determined below threshold. In fact, the deviation from zero of
eq.~\flavcomsnsb\ provides a simple way of testing for the presence of
intrinsic charm or anticharm, at least within the limitations of the
leading--order approximation.  The strange distribution can be
determined by comparing \eg\ eq.~\flavcomb\ below and above threshold.

Alternatively, above threshold, individual flavor contributions to the
structure functions can be separated by tagging the final-state quark:
\eg\ the strange contribution can be measured by tagging charm in the
final state, as is routinely done in unpolarized
experiments~\nudisrev. A simpler option consists in assuming that the
charm distribution vanishes below threshold, and is generated
dynamically by perturbative evolution above threshold. In such case,
as mentioned above, the up and down components can be determined from
eqs.~\flavcoma,\flavcomb\ below threshold, and the strange
distribution can then be determined from
eqs.~\flavcomsnsb,\flavcomsnsc\ above threshold.

As already mentioned, the virtual photon structure functions
eq.~\gnclo\ do not provide any further information. However, they do
provide independent consistency checks. Specifically, as is well
known, the isotriplet combination of $g_1^{\gamma^*}$ is particularly
interesting in that its first moment is determined by the Bjorken sum
rule. Now, the same isotriplet combination of parton distributions is
measured by neutral--current and charged--current structure functions:
\eqn\bjsr{6g_1^{\gamma^*}[p-n]=\left(g_5^{W^+}-g_5^{W^-}\right)[p-n]
=\left(\Delta
u+\Delta \bar u\right)- \left(\Delta d+\Delta\bar d\right).}
Likewise, the strange distribution can be also determined by combining
neutral-- and charged--current structure functions:
\eqn\strng{6g_1^{\gamma^*}[p+n]-\smallfrac{5}{3}
\left(g_1^{W^+}+g_1^{W^-}\right)[p]=
\left(g_5^{W^+}-g_5^{W^-}\right)[p+n]=\left(\Delta
c+\Delta \bar c\right)- \left(\Delta s+\Delta\bar s\right).}
Finally, notice that a determination of the pure
singlet quark distribution $\Delta \Sigma$ eq.\qsdef\ can be directly obtained 
combining charged--current $g_1$ data:
\eqn\sig{\left(g_1^{W^+}+g_1^{W^-}\right)[p]=\left(g_1^{W^+}+
g_1^{W^-}\right) [n]
=\Delta\Sigma^+.}
This is to be contrasted with the analogous measurement obtained from
global fits to $g_1^{\gamma^*}$ data, which can only be done (for
first moments) using information from octet $\beta$--decays and SU(3)
symmetry, or else (for all moments) by using scaling violations. A
potentially very accurate measurement of this quantity, whose
smallness is at the origin of the so-called `nucleon spin crisis', is
thus possible.

Of course, all combinations of parton distributions given in this
section are only correct at leading order, and may thus only be used
for rough estimates. Next-to-leading order corrections, which are
necessary for a more detailed treatment, would modify these relations.
In practice, an accurate determination of parton distributions can be
achieved by means of a global fit, which will include all these
relations and their next-to-leading order modifications.

\newsec{Structure functions and parton distributions at next-to-leading
order}

Beyond leading order, structure functions are no longer linear
combinations of quark distributions, rather, they are obtained by
convoluting quark and gluon distributions with perturbative
coefficient functions; furthermore, the splitting functions which
govern their evolution eq.~\apeq\ acquire a dependence on $Q^2$
through $\as$, and in particular no longer satisfy the simple
relations eqs.~\lopqq-\vanlop.

\subsec{Next-to-leading coefficient functions}

The general expression of the leading--twist polarized charged--current
structure functions beyond leading order is
\eqn\ggen{\eqalign{
g^{W^\pm}_1(x,Q^2)&={1\over2}\left[
    \Delta C_{\Sigma^+}\otimes\Delta\Sigma^+ 
\mp \Delta C_{NS\,-}\otimes\left(\Delta q^{NS\,-}_{ud}
+\Delta q^{NS\,-}_{cs}\right)
\right]+2[n_f/2] \Delta C_g\otimes\Delta g
\cr
g^{W^\pm}_i(x,Q^2)&=-{1\over2}\left[
    \Delta C_{\Sigma^-}^i\otimes\Delta\Sigma^- 
\mp \Delta C^i_{NS\,+}\otimes\left(\Delta q^{NS\,+}_{ud}
+\Delta q^{NS\,+}_{cs}\right)
\right],\,i=4,5,\cr}}
where $n_f$ is the number of active flavors, and $[n_f/2]$ is the
integer part of $n_f/2$.  In this equation, $\Delta g$ is the
polarized gluon distribution, the singlet quark distribution
$\Delta\Sigma^\pm$ and nonsinglet distributions $\Delta
q^{NS\,\pm}_{ij}$ were defined in eq.~\qsdef\ and eq.~\qnsdef\
respectively, and all coefficient functions admit a perturbative
expansion of the form $\Delta C(x,\as)=\Delta C^{(0)}+
{\as\over2\pi}C^{(1)}(x,\as)+\dots\>$, where $\Delta
C^{(0)}=\delta(1-x)$ for all the quark coefficient functions in
eq.~\ggen, and $\Delta C^{(0)}_g=0$ for the gluon coefficient
function.  Notice that the polarized gluon distribution contributes to
the structure function $g_1$, but decouples from $g_4$ and $g_5$ by
charge conjugation.

To next-to-leading order, in fact, for each of the three structure
functions  all quark coefficient functions are equal in standard
factorization schemes such as \MS:
\eqn\nlocf{\eqalign{
\Delta {C_{\Sigma^+}}^{(1)}(x)
&=\Delta {C_{NS\,-}}^{(1)}(x)\equiv \Delta {C_q}^{(1)}(x)\cr
\Delta {C_{\Sigma^-}^i}^{(1)}(x)
&=\Delta {C_{NS\,+}^i}^{(1)}(x)\equiv \Delta {C_i}^{(1)}(x)
,\quad i=4,5\cr}}
so that eq.~\ggen\ reduces to the simpler expression
\eqn\gnlo{\eqalign{g^{W^\pm;\,{\rm NLO}}_1(x,Q^2)&=\Delta C_q\otimes
g^{W^\pm,\,{\rm LO}}_1+2[n_f/2]\,\Delta C_g\otimes\Delta g\cr
g^{W^\pm,\,{\rm NLO}}_i(x,Q^2)&=\Delta C_{i}\otimes g^{W^\pm,\,{\rm LO}}_i,\quad i=4,5,\cr}}
where $g^{W^\pm;\,{\rm LO}}$ are the leading--order expressions
of the various structure functions, given by eq.~\gglo\ above charm
threshold and eq.~\gglobct\ below charm threshold.
Note, however, that the simple relations eqs.~\nlocf\ are already
violated at order $\alpha_s^2$ in the \MS\ scheme~\vn.

The next-to-leading coefficient functions for virtual photon DIS
structure functions have been known for some time~\koda. Because the
underlying partonic subprocesses are the same, the $g_1$ coefficient
functions for charged--current and neutral--current scattering in fact
coincide: the only difference is in the couplings between gauge bosons
and quarks, which is reflected in the different leading--order
expressions eq.~\gglo\ and eq.~\gnclo.  As is by now well
known~\spinrevth, the first moment of the singlet polarized quark
distributions has an O(1) scheme dependence, due to the fact that at
leading order the first moment $\Delta g(1)$ of the gluon distribution
evolves as $1/\as$, \ie\ $\as \Delta g(1)$ is scale independent up to
next-to-leading corrections. This is a consequence~\anom\ of the
anomaly which affects the singlet axial current.  A further
consequence of this is that it is possible to choose the factorization
scheme in such a way that the first moment of the singlet quark
distribution is scale independent (to all perturbative orders), even
though this is not the case in the \MS\ scheme. This choice is
particularly convenient in that it allows a meaningful comparison with
constituent quark model predictions and expectations, which are scale
independent. Even after fixing by this requirement the scheme
ambiguity on the quark first moment, there remains a residual
ambiguity on the other moments, which can be fixed in a minimal way by
requiring the scheme-change matrix from the \MS\ scheme to be
$x$-independent. This defines the so-called Adler-Bardeen (AB)
scheme~\bfr.

The first moment of any quark distribution in the \MS\ and AB schemes
are related by
\eqn\schch{\Delta q_i^{\rm\overline{MS}}
(1,Q^2)=\Delta q_i^{\rm AB}(1)- {\as\over 4\pi} \Delta g(1,Q^2),}
where $\Delta q(1,Q^2)=\int_0^1 dx \Delta q(x,Q^2)$.
Correspondingly, the first moment $\Delta C_g(1,\as)$ of the gluon
coefficient function vanishes in the \MS\ scheme, while in the AB
scheme it is equal to $\Delta C_g^{\rm AB} (1,\as)= -{\as\over 4\pi}$.
Notice that, below charm threshold, scheme invariance of the
physically observable structure function $g_1^{W^\pm}$ eq.~\ggen\ upon
this scheme change (and more generally upon any scheme change which
mixes the quark singlet and gluon distributions) is ensured by the
fact that the mixing with the gluon of $\Delta s$ and $\Delta d$
contributions combine to give one effective active flavor.  The
explicit expressions of the quark and gluon $g_1$ coefficient
functions are given \eg\ in Ref.~\bfr.

The next-to-leading order coefficient functions for the structure
functions $g_4$ and $g_5$ have been computed in Ref.~\deflo. Both in
the \MS\ and \AB\ schemes, they are simply related to the $g_1$ quark
coefficient function by
\eqn\cfourfiv{\eqalign{\Delta C_4^{(1)}(x)&=\Delta C_{q}^{(1)}(x)+C_Fx(1+x)\cr
\Delta C_5^{(1)}(x)&=\Delta C_{q}^{(1)}(x)+C_Fx(1-x),\cr}}
where $C_F={N^2_c-1\over 2N_c}$.  The fact that these coefficient
functions are unequal to each other implies in particular that the
Callan-Gross-like relation, eq.~\cgrel, is violated beyond leading
order, as expected since the helicity conservation on which it was
based only holds at leading order. In fact, the relations between the
three quark coefficient functions eq.~\cfourfiv\ are the same as the
corresponding relations between unpolarized quark coefficient
functions which are obtained by the replacements eq.~\polunpol, again
due to the fact that the corresponding tensor structures in
eq.~\hadten\ are the same. However, by charge conjugation, there is no
gluon contribution to either $g_4$ or $g_5$, so the violation of
eq.~\cgrel\ is entirely given by eq.~\cfourfiv, contrary to the
unpolarized case where the Callan-Gross relation is also violated by a
gluon contribution. Hence, unlike $F_1$ and $F_2$ which, beyond
leading order, measure different combinations of parton distributions
(so in particular their comparison allows a direct extraction of the
gluon distribution), $g_4$ and $g_5$ measure the same combination of
parton distributions, and thus their separate determination does not
provide any extra information. Therefore, we will henceforth only
discuss the structure function $g_5$, since $g_4$ can be entirely
determined from it. Once again, this fact can be tested directly on
the data, by performing a three-parameter (\ie\ $g_1$, $g_4$, and
$g_5$) fit of the $y$ distributions.

\subsec{Evolution equations}

Beyond leading order the structure of the evolution equations is
complicated~\guidorev\ by the fact that now both $P_{q \bar q}$ and
$P^{ND}_{q_iq_j}$, as well as their polarized counterparts $\Delta
P_{q\bar q}$ and $\Delta P^{ND}_{q_iq_j}$ no longer vanish: at
next-to-leading order they are all given by the diagram displayed in
Fig.~1.  Furthermore, already at NLO $P^{ND}_{q\bar q}\not= P^D_{q
q}$, because~\rosac\ there are two identical quarks in the final state
in the diagonal transition, but not in the non-diagonal one. However,
\eqn\nlondeq{P^{ND,\,{\rm NLO}}_{q\bar q}=P^{ND,\,{\rm NLO}}_{qq},}
because at this order both processes proceed through the diagram of
Fig.~1.
\topinsert
\vbox{
\epsfxsize=10truecm
\centerline{\epsfbox{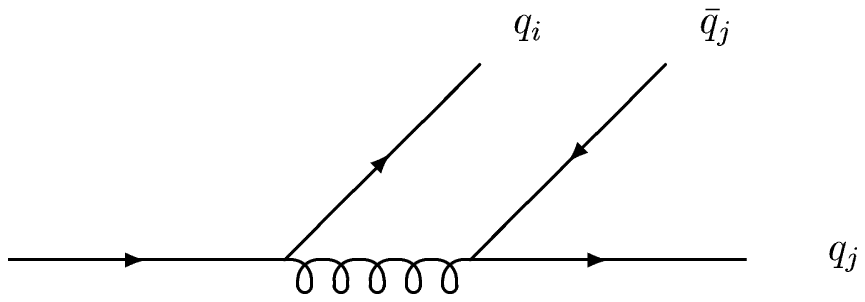}}
\vskip-1.truecm
\hskip4truecm\hbox{
\vbox{\footnotefont\baselineskip6pt\narrower\noindent
Figure 1: The lowest--order diagram which contributes to $P_{q\bar q}$ and
$P^{ND}_{q q}$.
}}\hskip4truecm}
\medskip
\endinsert

Coming to the polarized case, it is useful to recall that unpolarized
and polarized splitting functions are respectively~\ap\ the sum and
difference of helicity conserving and helicity flipping transitions:
\eqn\polunpolp{P=P^{\uparrow\uparrow}+P^{\uparrow\downarrow},\quad
\Delta P=P^{\uparrow\uparrow}-P^{\uparrow\downarrow} \, ,} 
where the arrows here refer to the helicities of the parton before and
after the splitting. Due to the collinearity of the splitting,
parallel (antiparallel) helicities are equivalent to parallel
(antiparallel) spins.  Because the quark-gluon coupling conserves
helicity, it follows that in the helicity-flipping transition there
are two identical particles in the final state (including the spin
quantum number), but not in the helicity conserving one. Therefore,
\eqn\poluneq{
P^{D\,\uparrow\downarrow}_{q\bar q}\not
=P^{ND\,\uparrow\downarrow}_{q\bar q},}
but not for the opposite spin configuration:
in fact, \eqnn\poleqa\eqnn\poleqb\eqnn\poleqc
$$\eqalignno{&
P^{D\,\uparrow\uparrow,\,{\rm NLO}}_{q\bar q}
=P^{ND\,\uparrow\uparrow,\,{\rm NLO}}_{q\bar q};&\poleqa\cr&
P^{ND\,\uparrow\uparrow,\,{\rm NLO}}_{q\bar q}=
P^{ND\,\uparrow\downarrow,\,{\rm NLO}}_{q\bar q}=
P^{ND\,\uparrow\uparrow,\,{\rm NLO}}_{qq}
=P^{ND\,\uparrow\downarrow,\,{\rm NLO}}_{qq};&\poleqb\cr&
P^{ND\,\uparrow\downarrow,\,{\rm NLO}}_{qq}=
P^{D\,\uparrow\downarrow,\,{\rm NLO}}_{qq},&\poleqc\cr}$$
where the second equation follows from the fact that the equality of
splitting functions eq.~\nlondeq\ holds for any spin configuration,
and the last equation is a consequence of the fact that, because of
helicity conservation, the flavor-diagonal but spin-flip splitting
also proceeds only through the diagram displayed in Fig.~1.

Now, eq.~\poluneq\ immediately implies that the C--even and C--odd
nonsinglet splitting functions are not the same: $\Delta
P_{NS}^+\not=\Delta P_{NS}^-$, like~\rosac\ their unpolarized
counterparts. However, using eq.~\poleqb\ in the explicit expression
for the nonsinglet splitting functions eq.~\pexp\ it immediately
follows that
\eqn\pmineq{
\Delta {P^{-,\,{\rm NLO}}_{NS}}=\Delta {P^{-,\,{\rm NLO}}_{S}}=
\Delta {P^{D,\,{\rm NLO}}_{qq}}-\Delta {P^{D,\,{\rm NLO}}_{q\bar q}}
\equiv
\Delta {P^{-,\,{\rm NLO}}}}
and similarly for their unpolarized counterparts (with $\Delta P\to P$).
Namely, at next-to-leading order there is only one C--odd
polarized (and one unpolarized) splitting function, and thus each
C--odd distribution $\Delta q^-_i$ eq.~\cevod\ 
for the $i$-th flavor evolves independently:
\eqn\pminev{\frac{d}{dt}\Delta q^-_i
=\frac{\as(t)}{2\pi} \Delta P^{-} \otimes \Delta q^-_i.}
The difference between singlet and nonsinglet C--odd splitting
functions starts at order $\as^2$, but is expected to be very
small~\nnlovvn\ at that order.

Therefore, at next-to-leading order the full set of evolution
equations is given by supplementing the standard singlet and
nonsinglet C--even evolution equations eq.~\apeq\ with the C--odd
evolution equation eq.~\pminev. The two-loop polarized splitting
functions have been computed recently~\nload, and are specifically
discussed in the AB and related factorization schemes in Ref.~\bfr.
The C--odd splitting function, and in fact all the nonsinglet
splitting functions~\nspol, can be determined from their unpolarized
counterparts. Indeed, using respectively eq.~\poleqa\ and \poleqc\ in
the definitions of eq.~\polunpol\ of $\Delta P^{NS}_{q\bar q}$ and
$\Delta P^{NS}_{q q}$ we get
\eqn\nsexp{\eqalign{&\Delta P^{NS,\,{\rm NLO}}_{q
q}=P^{NS\uparrow\downarrow,\,{\rm NLO}}_{qq} =P^{NS,\,{\rm
NLO}}_{qq}\cr
&\Delta P^{NS,\,{\rm NLO}}_{q\bar
q}=-P^{NS\uparrow\downarrow,\,{\rm NLO}}_{q\bar q} =P^{NS,\,{\rm
NLO}}_{q\bar q},\cr}}
which imply
\eqn\nspolunpol{\Delta P^{\pm,\,{\rm NLO}}_{NS}=P^{\mp,\,{\rm
NLO}}_{NS}.}
In Mellin space, the anomalous dimensions
\eqn\mel{\gamma_{NS}(N)\equiv\int_0^1 dx\,x^{N-1} P_{NS}(x)}
do not admit an analytic continuation for all $N$; however, their
even and odd moments can be analytically continued separately for all
$N$. Because even (odd) moments correspond to C--even (C--odd)
operators, the respective analytic continuations give the moments of
$P^+_{NS}=\Delta P^-_{NS}$ and $P^-_{NS}=\Delta P^+_{NS}$~\rosac.
Equations~\poleqa-\poleqc\ and their consequences hold in the \MS\
scheme, and remain valid in schemes, such as the AB scheme~\bfr\ where
the nonsinglet and C--odd quark anomalous dimensions are the same as
in \MS.

\newsec{Theoretical expectations and constraints}

Even though neutral--current structure function data do not give any
information on the relative size of quark and antiquark (or `valence'
and `sea') distributions, some experimental information is available
from semi-inclusive polarized experiments~\semexp. However, the only
conclusion which can be drawn from these experiments at present
is that the first
moments of the $\Delta \bar u(x)$ and $\Delta 
\bar d(x)$ distributions are much smaller
than the C--even combinations $\Delta u(x)+\Delta \bar u(x)$ and
$\Delta d(x)+\Delta  \bar d(x)$, and in fact
compatible with zero. When the same data are folded in a global NLO
analysis~\semifit\ they  lead to an indication that the first
moment of the $\Delta \bar
u(x)$ distribution is positive, and roughly 10\% of the first moment
of $\Delta u(x)+ \Delta \bar
u(x)$, but since this is smaller than the typical systematic
uncertainties on polarized first moments~\cracovia\ the only solid
conclusion that can be drawn from this is, again, that the light antiquark
distributions are significantly smaller than the C--even combinations
measured in inclusive experiments. Therefore, a wide variety of
theoretical scenarios is compatible with present-day data. However,
even in the absence of experimental information, some constraints on
the relative size of the $\Delta  q$ and $\Delta \bar q$ distributions
are imposed by the requirement of positivity of physical cross--sections.

\subsec{Positivity bounds}

Positivity bounds~\pos\ on polarized structure functions follow from
their definition in terms of a difference of cross--sections
eq.~\xsecas, while their unpolarized counterparts are defined from the
sum of the same cross--sections. Expressing structure functions in
terms of parton distributions then leads to bounds on the parton
distributions themselves. If the leading--order expressions of
structure functions are used, these bounds coincide with those which
are obtained in the naive parton model by interpreting parton
densities as (positive-semidefinite) number densities, while beyond
leading order they differ from the naive partonic bounds by a
calculable amount.  Of course, given accurate measurements of all the
relevant quantities, positivity will take care of itself, \ie\ it will
just be a trivial consequence of the fact that cross--sections are
positive. However, in the presence of incomplete information,
positivity imposes nontrivial bounds on the possible outcomes of
future experiments.

In Ref.~\pos, positivity bounds on the C--even quark distributions and
the gluon were derived from their definitions in terms of virtual
photon DIS, while it was mentioned (but not proven) that
charged--current DIS implies similar bounds on quark and antiquark
distributions separately. In order to prove such bounds, it is enough
to consider the cross--section asymmetry for gauge boson-hadron
scattering $A_1$, analogous to that discussed in Ref.~\pos, but with
the $\gamma^*$ beam replaced by a $W^\pm$ beam. To this purpose,
consider the cross--section for scattering of a $W$ with helicity
$\lambda_W$ off a hadron with helicity $\lambda_p$, given by
\eqn\sigbosh{\sigma^{W^\pm}(\lambda_W,\lambda_p)= K
{\epsilon^\mu}(\lambda_W){\epsilon^\nu}^*(\lambda_W)
W_{\mu\nu}^{W^\pm}(\lambda_p),} where $K$ contains the coupling
constants and the flux and phase-space factors. The polarization
vector is $\epsilon^\mu(\pm 1)=\smallfrac{1}{\sqrt{2}}(0,1,\pm i,0)$.

A straightforward calculation using the definition eq.~\hadten\ of the
hadronic tensor leads to
\eqn\helsig{\sigma^{W^\pm}(\lambda_W,\lambda_p)
=K\left[\left(F_1^{W^\pm}-\lambda_W F_3^{W^\pm}/2\right)
+\lambda_p\left(g_5^{W^\pm}+\lambda_W g^{W^\pm}_1\right)\right].}
Therefore, we get
\eqn\asymdef{\eqalign{
A_1^{W^\pm}(\lambda_W,x,Q^2)
&\equiv{\sigma(\lambda_W,-1)-\sigma(\lambda_W,+1)
\over
\sigma(\lambda_W,-1)+\sigma(\lambda_W,+1)}
\cr&=-{g^{W^\pm}_5(x,Q^2)+\lambda_W g^{W^\pm}_1(x,Q^2)
\over F_1^{W^\pm}(x,Q^2)-\lambda_W F^{W^\pm}_3(x,Q^2)/2}.\cr
}}
The asymmetry eq.~\asymdef\ is equal to half (because of the
factor discussed after eq.~\polunpol) the ratio of the cross-sections
eq.~\xsecas\ and \xsectot\ evaluated at $y=1$. This is a consequence
of the fact that when $y=1$ the lepton and the gauge boson are
collinear, and their helicities must then be aligned by helicity and
momentum conservation, so $\lambda_W=\lambda_\ell$.

The leading order positivity bounds now follow immediately by noticing
that, because of eq.~\gglo\ and its unpolarized counterpart, the four
combinations in the numerator and denominator correspond respectively
to the sum over generations of the two unpolarized and polarized
flavors and antiflavors in each generation: \eg, above top threshold
\eqn\explas{|A_1^{W^+,\,{\rm LO}}(-1,x,Q^2)|
={\left|\Delta d(x,Q^2)+ \Delta s(x,Q^2)+\Delta
b(x,Q^2)\right|\over d(x,Q^2)+ s(x,Q^2)+ b(x,Q^2)}\leq 1.} 
The three other combinations give the analogous expressions for
$u$--type quarks and for the $u$--type and $d$--type antiquarks.
Because the bound $|A_1|\le 1$ must be satisfied as a matter of
principle for any choice of target, below and above each threshold,
and also for the process where the flavor of the final--state quark is
tagged, we immediatley get the bounds of Ref.~\pos
\eqn\posboun{|\Delta q_i(x,Q^2)|\le q_i(x,Q^2);\qquad|\Delta \bar
q_i(x,Q^2)|\le \bar q_i(x,Q^2)} for any flavor $i$.  It is easy to see
that these conditions are sufficient to ensure positivity of the
physical lepton--hadron cross--section. Indeed, using the
leading--order expression of the structure functions eq.~\gglo\ in the
cross--section eq.~\xsectot,\xsecas, the positivity conditions for the
$\nu$--hadron cross section is \eqn\poshad{|\Delta \bar u -
(y-1)^2\Delta d |\leq\bar u - (y-1)^2 d ,} which is manifestly true if
eq.~\posboun\ holds. Similar conditions hold for the other choices of
lepton beam.

As discussed in Ref.~\pos, the Altarelli-Parisi equations~\apeq\ imply
that if the positivity bounds are respected at some scale $Q_0^2$ by a
given set of polarized and unpolarized parton distributions, they will
also be respected by the same parton distributions for all $Q^2>
Q_0^2$. Conversely, this implies that the bounds will always be
violated at sufficiently low scale. However, at low scale the
leading--order approximation breaks down, and it is necessary to
consider higher-order corrections. When the next-to-leading order
expressions of the structure functions eq.~\gnlo\ are used, the
positivity bounds are best written in terms of the Mellin transforms
(defined as in eq.~\mel) of the coefficient functions and parton
distributions, so that all convolutions turn into ordinary
products. In such case, the next-to-leading order bound eq.~\explas,
including for simplicity only the lightest flavor (as one gets below
charm threshold neglecting Cabibbo-suppressed contribution) becomes
\eqn\explasnl{
{\left|\left[1+\smallfrac{\as}{4\pi}\left(\Delta C_q^{(1)}+\Delta
C_5^{(1)}\right)
\right]\Delta d+\smallfrac{\as}{4\pi}
\left(\Delta C_q^{(1)}-\Delta C_5^{(1)}\right) \Delta \bar
u+\smallfrac{\as}{2\pi}\Delta  C_g^{(1)}\Delta g\right|\over
\left[1+\smallfrac{\as}{4\pi}\left(C_q^{(1)}+C_3^{q\,(1)}\right)
\right]d+\smallfrac{\as}{4\pi}
\left( C_q^{(1)}- C_3^{(1)}\right) \bar
u+\smallfrac{\as}{2\pi} C_g^{(1)} g}\leq1,}
where $C_q$, $C_g$ and $C_3$ are the unpolarized $F_1$ and $F_3$
coefficient functions, and all parton distributions are functions of
$Q^2$ and the Mellin variable $N$.

\topinsert
\vbox{
\epsfxsize=10truecm
\centerline{\epsfbox{fig2.ps}}
\vskip-.4truecm
\hskip4truecm\hbox{
\vbox{\footnotefont\baselineskip6pt\narrower\noindent
Figure 2: Leading--order positivity bounds for the C--odd up distribution
(solid). The $\Delta\bar{u}=0$ curve (dashed) is also shown.
}}\hskip4truecm}
\medskip
\endinsert
The impact of the gluon correction on the quark positivity bounds has
been extensively discussed in Ref.~\pos. Because gluons only mix with
the C--even combination of quark distributions, and thus enter
eq.~\explasnl\ through the $g_1$ contribution, the conclusions of that
reference are unchanged. Namely, because the gluon distribution is
peaked at small $x$, the correction to the leading--order positivity
bound is only sizable for small moments $N\sim 1.5$, but there the
bound itself is very loose because $g(N)$ and $q(N)$ diverge as
$N\to 1$ while $\Delta g(N)$ and $\Delta q(N)$ remain finite, so there
is no bound as $N\to 1$. Thus, unless the scale is outside the
perturbative region (for instance $Q^2\lsim 1$~GeV$^2$), gluon-driven
next--to--leading order corrections to the positivity bound are
essentially negligible. On top of gluon corrections, eq.~\explasnl\
also contains a correction due to the fact that the quark coefficient
functions for $g_1$ and $g_5$ are not the same, as per eq.~\cfourfiv.
This difference, however, is actually quite small and rapidly
decreasing at large $x$: indeed, from eq.~\cfourfiv\ we get that
$\Delta C^{(1)}_5(N)-\Delta C^{(1)}_q(N)={1\over (N+1)(N+2)}$. Hence
even at $Q^2\sim 1$~GeV$^2$ where $\as\sim 0.5$, and even for,
say, $N=1.5$ the correction is around $1\%$, and rapidly decreasing as
$N$ increases: but, again, at low $N$ the bound is not relevant
because of the vanishing of $\Delta q_i(N)\over q_i(N)$ as $N\to 1$.
Hence, to very good approximation, the next-to-leading order
positivity bounds are
\eqn\approxnlob{\left|{\Delta q(N)\over
q(N)}\right|\leq {1+\smallfrac{\as}{2\pi}C^{(1)}_q(N)\over
1+\smallfrac{\as}{2\pi}\Delta C^{(1)}(N)},}
\ie\ the only significant correction to the leading--order bound is
due to the fact that the polarized and unpolarized quark coefficient
functions are unequal to NLO. Because however
$\lim_{N\to\infty}{C^{(1)}_q(N)\over\Delta C^{(1)}(N)}=1$, this
difference is again only significant if $N$ is not too large: in
practice, the correction was shown in Ref.~\pos\ to lead to a
modification of a few percent of the leading--order bound for moments
around $N\sim 5$, where positivity bounds can be relevant.  We
conclude that for all practical purposes it is enough at the present
stage to consider leading--order positivity bounds, since
next--to--leading corrections, in the form of eq.~\approxnlob, are
only relevant at the level of accuracy of a few percent, and then only
for a limited range of moments, relevant for the description of the
shape of the parton distributions when $x\gsim 0.1$.

\topinsert
\vbox{
\epsfxsize=10truecm
\centerline{\epsfbox{fig3.ps}}
\vskip-.4truecm
\hskip4truecm\hbox{
\vbox{\footnotefont\baselineskip6pt\narrower\noindent
Figure 3: Leading--order positivity bounds for the C--odd down distribution
(solid). The $\Delta\bar{d}=0$ curve (dashed) is also shown.
}}\hskip4truecm}
\medskip
\endinsert
Using the experimentally measured values of the unpolarized parton
distributions and the polarized C--even distributions, we can then turn
the pair of positivity bounds eq.~\posboun\ on quark and antiquark
distributions for each flavor into a bound on the admissible value of
the unknown C--odd quark distribution $\Delta q_i^-$ eq.~\cevod.  The
uncertainty on the bound will be dominated by, and thus essentially of
the same size as, that on the polarized quark distributions~\smcfit,
which are much more poorly known than the unpolarized ones.  In
Figs.~2-4, we show the positivity bounds on the up, down and strange
distributions at $Q^2=5$~GeV$^2$ obtained in this way by using the
CTEQ5~\cteq\ unpolarized up and down quark and antiquark distributions
and the BPZ~\bpz\ strange distribution, and the polarized parton
distributions from Ref.~\cracovia. Use of the BPZ fit is motivated by
the fact that in this fit a particularly accurate determination of the
unpolarized strange quark and antiquark distributions was achieved by
means of a detailed analysis of available neutrino DIS data, whereas a
separate determination of the strange quark is not attempted by the
CTEQ group (which only determines an SU(3) symmetric sea with
corrections for the $\bar u-\bar d$ asymmetry).  In fact, it turns out
that the C--even polarized distribution from Ref.~\cracovia\ is
incompatible with the leading--order positivity bound obtained from
the CTEQ5 unpolarized strange distribution at large $x$
($x\gsim0.5$). Even though the violation is within theoretical errors,
it is suggestive that no such violation is found when the nominally
more accurate BPZ strange distribution is used.

\topinsert
\vbox{
\epsfxsize=10truecm
\centerline{\epsfbox{fig4.ps}}
\vskip-.4truecm
\hskip4truecm\hbox{
\vbox{\footnotefont\baselineskip6pt\narrower\noindent
Figure 4: Leading--order positivity bounds for the C--odd strange distribution
(solid). The $\Delta\bar{s}=0$ curve (dashed) is also shown.
}}\hskip4truecm}
\medskip
\endinsert
In Figs.~2-4, we also display the $\Delta q_i^-=\Delta q_i^+$, \ie\
$\Delta \bar q_i=0$ curve for each flavor (dashed curves).  It is
clear that a large value of the antiquark distribution is not
compatible with positivity for both the up and down quark
distributions. Indeed, the curve $\Delta \bar q= \Delta q$ (\ie\
$\Delta q^-=0$) violates the positivity bound for the up quark when
$x\gsim 0.1$ and for the down quark when $x\gsim 0.2$, while when
$x\approx 0.3$ only a very small value of the polarized antiquark
distribution is allowed (about an order of magnitude smaller than the
polarized quark distribution). On the other hand, a vanishing light
antiquark distribution is compatible with positivity for all $x$.  In
the case of the strange distribution, instead, both $\Delta \bar s=0$
and $\Delta \bar s=\Delta s$ (or indeed $\Delta \bar s=-\Delta s$) are
compatible with positivity.  As one should expect, for the strange
polarized distribution, positivity does not force the antiquark to be
smaller than the quark distribution.

\subsec{Theoretical scenarios and the spin of the proton}

As we have already mentioned in the introduction, one of the main
reasons of interest in polarized quark distributions is the unexpected
smallness of the nucleon axial charge which has been determined in the
first generation of polarized DIS experiments. A clarification of the
physics behind this requires a determination of the detailed polarized
parton content of the nucleon. A review of the various explanations
which have been proposed for this experimental fact is beyond the
scope of this work. However, it is useful to sketch various scenarios
for the polarized content of the nucleon which are representative of
possible theoretical alternatives, and which could be tested in future
experiments.

Firstly, it should be noticed that even though current data give a
value of the axial charge which is compatible with zero, they cannot
exclude a value as large as $a_0(10$~GeV$^2)=0.3$~\cracovia. Also, the
current value is obtained by using information from hyperon
$\beta$--decays  and
SU(3) symmetry. Clearly, the theoretical implications of an exact zero
are quite different from those of a value which is just smaller than
expected in quark models. It is thus important to have a direct
determination of the axial charge. If a small value 
is confirmed, it could be understood as the consequence of a
cancellation between a large value of the scale--independent (\ie\
AB--scheme) quark first moment, and a large gluon first moment.
Indeed, since in the \MS\ scheme the axial charge and singlet quark
first moment coincide~\anom, eq.~\schch\ is immediately seen to imply
that a large gluon contribution can lead to a small value of the axial
charge even when the AB--scheme quark is large.  In this (`anomaly')
scenario the up, down and strange polarized distributions in the
AB--scheme are close to their expected quark model values, so in
particular the strange distribution is much smaller than the up and
down distributions.  In Ref.~\topsup, this cancellation of quark and
gluon components has been derived from the topological properties of
the QCD vacuum~\topsup\ (and thus further predicted to be a universal
property of all hadrons).

If instead the polarized gluon distribution is small, the smallness of
the singlet axial charge can only be explained with a large and
negative strange distribution. In this case, the scale--independent
first moment of the singlet quark distribution is also small. This
scale--independent suppression of the axial charge might be explained
by invoking non--perturbative mechanisms based on instanton--like
vacuum configuration~\inst. In this `instanton' scenario the strange
polarized distribution is large and equal to the anti-strange
distribution, since gluon-induced contributions must come in
quark--antiquark pairs.

Another scenario is possible, where the smallness of the singlet axial
charge is due to intrinsic strangeness, \ie\ the C-even strange
combination is large, but the sizes of $\Delta s$ and $\Delta \bar s$
differ significantly from each other. Specifically, it has been
suggested that while the strange distribution (and specifically its
first moment) is large, the antistrange distribution is much smaller,
and does not significantly contribute to the nucleon axial
charge~\ints. This way of understanding the nucleon spin structure is
compatible with Skyrme models of the nucleon, and thus we will refer
to this as a `skyrmion' scenario~\strange.

Therefore, the main qualitative issues which are relevant for the
nucleon spin structure are to assess how small the axial charge is, to
determine whether the polarized gluon distribution is large, and then
whether the strange polarized distribution is large, and whether the
strange polarized quark and antiquark distributions are equal to each
other or not.  More detailed scenarios might then be considered, once
the individual quark and antiquark distributions have all been
accurately determined. For instance, while the up and down antiquark
distributions are small, they need not be zero, and in fact they could
be different from each other~\glumod, just like their unpolarized
counterparts appear to be. Investigating these issues could shed
further light on the detailed structure of polarized nucleons.

\newsec{Polarized neutrino-DIS at the front-end of a muon ring}

The parameters of a realistic neutrino-factory complex are still under
active study, and it is expected that the accelerator configuration
will evolve and upgrade in both energy and luminosity after the
beginning of operations~\nufactzz.  For the present study we shall
assume the following beam parameters, which are considered as typical
of the initial operations of a neutrino factory: muon beam energy,
$E_{\mu}=50$~GeV; length of the straight section, $L=100$~meters;
distance of the detector from the end of the straight section,
$d=30$~meters; number of muon decays per year along the straight
section, $N_{\mu}=10^{20}$; muon beam angular divergence, $0.1\times
m_{\mu}/E_{\mu}$; muon beam transverse size
$\sigma_x=\sigma_y=1.2$~mm.  While the concept of ``conservative''
cannot be applied to any of the projections for the neutrino factory,
it is nevertheless reasonable to expect that should one ever be built,
its performance would not be inferior to what assumed here.  We also
assume a cylindrical detector, with azimuthal symmetry around the beam
axis, parametrized by its radius $R=50$~cm and a thickness of $10~{\rm
gr}/{\rm cm}^2$.  The collected statistics scales linearly with the
detector thickness, while the dependence on other parameters such as the
radius or the length of the straight section is clearly more complex.

The neutrino spectra are calculated using standard expressions for the
muon decays (see \eg\ Ref.~\bkbook).  For simplicity we shall confine
ourselves to the case of $\nu_{\mu}$ and $\bar{\nu}_{\mu}$ charged
current DIS.  The laboratory-frame neutrino spectra, convoluted with
the charged--current interaction cross-sections, are shown for several
detector and beam configurations in Fig.~5 ($E_{\mu}=50$~GeV) and in
Fig.~6 ($E_{\mu}=100$~GeV).
\topinsert
\vbox{
\epsfxsize=13truecm
\centerline{\epsfbox{50gev.ps}}
\vskip-1.truecm
\bigskip
\hskip4truecm\hbox{
\vbox{\footnotefont\baselineskip6pt\narrower\noindent
Figure 5: Charged--current event rates, in units of $10^6$, as a
  function of Lab-frame neutrino spectra, for several detector and
  beam configurations. The dashed lines on the left include cuts on
  the final-state muon ($E_{\mu}>3$~GeV) and on the final-state
  hadronic energy ($E_{had}>1$~GeV). The solid lines have no
  energy-threshold cuts applied. The three set of curves correspond to
  different detector radii (50, 10 and 5~cm, from top to bottom).
}}\hskip4truecm}
\medskip
\endinsert

\topinsert
\vbox{
\epsfxsize=13truecm
\centerline{\epsfbox{100gev.ps}}
\vskip-1.truecm
\bigskip
\hskip4truecm\hbox{
\vbox{\footnotefont\baselineskip6pt\narrower\noindent
Figure 6: Same as Fig.~5, but for 100~GeV muon beams.
}}\hskip4truecm}
\medskip
\endinsert

To determine the statistical accuracy with which the individual
structure functions and their flavor components are expected to be
measured we shall exploit the different $y$ dependence of the separate
$F_{i}$ and $g_{i}$ components of the cross-section. An important
feature of the neutrino beams from muon decays is their wide-band
nature. This allows to modulate the $y$ dependence for fixed values of
$x$ and $Q^2$ using the neutrino energy: \eqn\ydep{y=\frac{Q^2}{2xm
E_{\nu}}.}  The separate measurement of the recoil muon energy and
direction, and of the recoil hadronic energy, enable the
event-by-event extraction of $x,y$ and $Q^2$. We assume perfect
experimental resolution on the determination of these quantities,
after imposing a cut on the minimum energy of the muon ($>3$~GeV) and
of the hadronic system ($>1$~GeV).  Since we are just interested in
predicting the statistical accuracy in the determination of the
structure functions, we also assume for simplicity the absence of
power-suppressed corrections, and impose the Callan-Gross relation.
Given the large statistics expected at the neutrino factory, it will
nevertheless be possible to use the data to extract the higher-twist
components and to separate the individual contributions of $F_1$ and
$F_2$ ($g_4$ and $g_5$), without any theoretical assumption. This
issue is studied in some detail for the case of unpolarized
distributions in Ref.~\cernrep.

We thus produced $y$ distributions by generating events within
different bins of $x$ and $Q^2$, and performed minimum-$\chi^2$ fits
of the generated data using the cross-section eq.~\xsectot. For each
bin, the values of $x$ and $Q^2$ at which we quote the results are
obtained from the weighted average of the event rate. As an input, we
used the CTEQ4D set of parton distributions~\Lai. The dependence on
the parameterization of the parton distributions is very small, and
will be neglected here. We verified that other recent sets of parton
distributions give similar results. The fit to the $y$ distributions
at fixed $x$ and $Q^2$ for a fully polarized target gives then the
value of the combinations $F_2\pm 2xg_5$ and $F_3\pm 2g_1$.

The absolute number of events expected in each bin is scaled by the
total number of muon decays; this number of events determines the
statistical error on the individual structure functions obtained
through the fit.  Polarization asymmetries are extracted by combining
data sets obtained using targets with different orientations of the
polarization. The statistical accuracies with which the combinations
can be performed depend on the statistical content of each individual
data set.  Since the polarization asymmetries are small relative to
the unpolarized cross-sections, the {\it absolute} statistical
uncertainties on the extraction of polarized structure functions will
have a very mild dependence on the value of the polarized structure
functions themselves; they will be mostly determined by the value of
the unpolarized structure functions (which to first approximation fix
the overall event rate), and by the polarization properties of the
target.

For simplicity, we therefore calculate directly the expected
statistical errors $\sigma_{F_2,F_3}$ on the extraction of $F_2$ and
$F_3$ using unpolarized targets, and relate them to the errors on the
polarized cross sections using the following relation given
in~\ballrev: \eqn\eqpolerr{ \sigma_{\rm g_i} = F^{tgt}_{\nu ,\bar\nu}
\sqrt{2} \; \alpha_{ij} \; \frac{\sigma_{\rm F_j}}{2} ,} where
$\alpha_{ij}=1$ for $(i,j)=(1,3)$ and $\alpha_{ij}=1/x$ for
$(i,j)=(5,2)$, and where $F^{tgt}_{\nu ,\bar\nu}$ is a correction
factor (always larger than one) which accounts for the ratio of the
target densities to $H_2$ or $D_2$, for the incomplete target
polarization, and for the dilution factor of the target, namely the
$\nu$ (or $\bar\nu$) cross-section weighted ratio of the polarized
nucleon to total nucleon content of the target.  The factor of
$\sqrt{2}$ in the numerator reflects the need to subtract the
measurements with opposite target polarization.

\topinsert
\vbox{
\epsfxsize=10truecm
\centerline{\epsfbox{xqrange.ps}}
\vskip-1.truecm
\bigskip
\hskip4truecm\hbox{
\vbox{\footnotefont\baselineskip6pt\narrower\noindent
Figure 7: $x$ and $Q^2$ binning for the generation of charged--current
events. The crosses correspond to the weighted bin centers.
}}\hskip4truecm}
\medskip
\endinsert

\topinsert
\vbox{
\epsfxsize=10truecm
\centerline{\epsfbox{perr.ps}}
\vskip-1.truecm
\bigskip
\hskip4truecm\hbox{
\vbox{\footnotefont\baselineskip6pt\narrower\noindent
Figure 8: Absolute errors on proton structure functions for the bins of Fig.~7.
}}\hskip4truecm}
\medskip
\endinsert
We generate events in the ($x,Q^2$) bins shown in Fig.~7.  Twenty
equally-spaced bins in the $0\leq y\leq 1$ range are used for the $y$
fit. The total number of $x$ bins varies in different $Q^2$ bins
because of kinematic acceptance and minimum energy cuts.  The values
of the uncertainties in the determination of the eight
charged--current structure functions ($g_1$ and $g_5$ with the two
available beams and targets) are assigned at the cross-section
weighted bin centers.  We then get the absolute errors on the
structure functions displayed in Figs.~8-9 for proton and deuterium
targets respectively, using the p-butanol and D-butanol
target~\smctarg\ correction factors given in Ref.~\ballrev, namely
$F^{p}_{\nu}=2.6$, $F^{p}_{\bar\nu}=1.6$ and $F^{D}_{\nu,\bar\nu}=4.4$
(for a more complete discussion of polarized targets and their
complementary properties, see~\ballrev\ and references therein).  We
have assumed a luminosity of $10^{20}$ muons decaying in the straight
section of the muon ring for each charge, for each target, and for
each polarization. Assuming that only one polarization and one target
can run at the same time, this means eight years of run. While the
number of muons may not be dramatically increased, the integration
time can be reduced by a large factor if the target thickness can be
increased over the conservative 10~gr/cm$^2$ assumed here, or if
different targets can be run simultaneously.  \topinsert \vbox{
\epsfxsize=10truecm \centerline{\epsfbox{derr.ps}} \vskip-1.truecm
\bigskip \hskip4truecm\hbox{
\vbox{\footnotefont\baselineskip6pt\narrower\noindent Figure 9:
Absolute errors on deuteron structure functions for the bins of
Fig.~7.  }}\hskip4truecm} \medskip 
\endinsert


\newsec{Determination of polarized structure functions at a neutrino factory}

We can now study how charged--current DIS data may be used to
determine the polarized parton content of the nucleon. The purpose of
this exercise is twofold. First, this will give us a first indication
of the expected shape and scale--dependence of the charged--current
structure functions. Furthermore, on a more quantitative level, we
will be able to assess the level of accuracy with which
charged--current DIS experiments will allow a determination of
individual flavors and antiflavors. Of course, both these tasks can
only be achieved by making assumptions on the expected flavor content
of the nucleon, and on the expected experimental accuracy. To this
purpose, we consider the theoretical scenarios which we have discussed
in Sect.~4, and we assume the availability of data such as discussed
in Sect.~5.

In order to implement the theoretical scenarios of Sect.~4, we need to
take as a starting point a set of parton distributions which summarize
the current knowledge of the polarized structure of the nucleon. To
this purpose, we adopt the type--A fit of Ref.~\cracovia, which is
defined as follows. The C--even polarized quark distributions
$\Delta\Sigma^+$ and
\eqn\nonsingdef{\eqalign{
\Delta q_3&\equiv\Delta q^{NS+}_{ud}\cr
\Delta q_8&\equiv\Delta q^{NS+}_{us}+\Delta q^{NS+}_{ds},\cr}}
and the polarized gluon distribution $\Delta g$ at the initial scale
$Q_0^2=1$~GeV$^2$ are all taken to be of the form
\eqn\fitashape{
\Delta f(x,Q_0^2)=
{\cal N}_f\eta_f x^{\alpha_f}(1-x)^{\beta_f}(1+\gamma_f x^{\delta_f}),}
where the factor ${\cal N}_f$ is such that the parameter $\eta_f$ is the
first moment of $\Delta f$ at the initial scale. The nonsinglet
quark distributions $\Delta q_3$ and $\Delta q_8$ are assumed to
have the same $x$ dependence, while
the parameter $\eta_8$, corresponding to the first moment of $\Delta q_8$,
is fixed to the value $\eta_8=0.579$ from octet
baryon decay rates using SU(3) symmetry.
Furthermore,
$\gamma_\Sigma=\gamma_g=10$, $\delta_3=\delta_8=0.75$, 
$\delta_\Sigma=\delta_g=1$. All other parameters in eq.~\fitashape\
are determined by the fitting procedure.

This set of polarized parton distributions is particularly useful for
our present purposes because a detailed analysis of statistical and
systematic uncertainties on the first moments based on it was
performed in Ref.~\cracovia. The set A is adopted because all parton
distributions in this set are amply within positivity
bounds~\pos. However, some new data~\newdata\ from the E155
collaboration, and the final data set from the SMC collaboration have
appeared since the publication of Ref.~\cracovia. Therefore, we have
repeated the fit including these new data, in order to ensure that no
available information is neglected. This gives us a data set of 176
neutral--current data points.

The best--fit values of the first moments of the C--even parton
distributions at the initial low scale $Q_0^2=1$~GeV$^2$ thus obtained
are listed in the first column of Table~1, together with the errors
from the fitting procedure. The fit is performed, and all results are
given, in the AB scheme. These values are not far from those of
Ref.~\cracovia. They are further affected by theoretical
uncertainties, for which we refer to~\cracovia.  The detailed
shapes of parton distributions are similar to those of Ref.~\cracovia,
and likewise not very accurately known.  As mentioned above, the value
of $\eta_8$ in this fit is fixed using the SU(3) flavor symmetry and
the measured octet baryon decay constants. Even though the nominal
uncertainty on this SU(3) value of $\eta_8$ is relatively small
(around 5\%), the uncertainty due to failure of exact SU(3) symmetry
could be as large as 30\%~\sutr. This effect is taken into
account in the estimated theoretical uncertainties given in
Ref.~\cracovia.  The subsequent three rows of the table give the
values of the first moments of $\Delta q(x,Q^2)+\Delta{\bar q}(x,Q^2)$
at $Q^2=1$~GeV$^2$ for up, down, and strange, obtained combining the
singlet and nonsinglet quark first moments above.  Finally, we give in
the last row the value of the singlet axial charge $a_0$ at the scale
$Q^2=10$~GeV$^2$.

Because the first moment of the gluon distribution in this fit is
quite large, we can take this global fit as representative of the
`anomaly' scenario discussed in Sect.~4.2, even though the strange
distribution is not quite zero.  In order to construct parton
distributions corresponding to the other scenarios discussed in
Sect.~4.2, we have also repeated this fit with the gluon distribution
forced to vanish at the initial scale. This possibility is in fact
disfavored by several standard deviations; however, once theoretical
uncertainties are taken into account a vanishing gluon distribution
can only be excluded at about two standard deviations~\cracovia, and
thus this possibility cannot be ruled out on the basis of present
data. The results of this fit for the various first moments are
displayed in the second column of Table~1.

We can now use these parton distributions to construct the unknown
C--odd parton distributions. We construct three sets of parton
distributions, corresponding to the three scenarios of Sect.~4.2. In
all cases, we assume $\Delta\bar u(x)=\Delta \bar
d(x)=0$. Furthermore, as the `anomaly' set we take the `generic' fit
of Table~1 with the assumption $\Delta \bar s(x)=0$, the strange
distribution for this set being relatively small anyway. As
`instanton' and `skyrmion' parton sets we take the $\Delta g=0$ fit of
Table 1, with $\Delta s=\Delta\bar s$ in the former case, and $\Delta
\bar s=0$ in the latter case. With these choices all quark and
antiquark distributions are fixed, and thus all structure functions
can be computed.

We generate for each of these three scenarios a set of pseudo-data, by
assuming the availability of neutrino and antineutrino beams, and
proton and deuteron targets, in the ($x,Q^2$) bins of Fig.~7 and with
the errors displayed in Figs.~8,9.  Although experimentally event rates
remain sizable even in the large--$x$ region, we do not include data
with $x>0.7$ because, in this region, leading twist next-to-leading
order perturbation theory is not reliable.  We discard data points
whose uncertainty is larger than 50, the typical size of the structure
functions being of order one. We finally use these uncertainties to
generate data gaussianly distributed about the values of the structure
functions at each data point in the three scenarios.  We obtain in
this way approximately 70 data points for each of the eight
charged--current structure functions.

\topinsert
\hbox{\vbox{\tabskip=0pt \offinterlineskip
      \def\tablerule{\noalign{\hrule}}
      \halign to 392pt{\strut#&\vrule#\tabskip=.5em plus2em
                   &#\hfil &\vrule#
                   &\hfil#&\vrule#
                   &\hfil#&\vrule#
                   &\hfil#&\vrule#
                   &\hfil#&\vrule#
                  &\hfil#&\vrule#\tabskip=0pt\cr\tablerule
             &&\omit par. \hidewidth
             &&\omit\hidewidth generic fit\hidewidth
             &&\omit\hidewidth $\Delta g=0$ fit\hidewidth
             &&\omit\hidewidth `anomaly' refit\hidewidth
             &&\omit\hidewidth `instanton' refit \hidewidth
             &&\omit\hidewidth `skyrmion' refit \hidewidth&\cr\tablerule
&& $\eta_\Sigma$&& $0.38\pm 0.03$ &&$0.31\pm 0.01$  &&$0.39\pm 0.01$   &&$0.321\pm 0.006$ && $0.324\pm 0.008$&\cr
&& $\eta_g$&& $0.79\pm 0.19$ &&$0$             &&$0.86\pm 0.10$   && $0.20\pm 0.06$  && $0.24\pm 0.08$  &\cr
&& $\eta_3$&&$1.110\pm 0.043$&&$1.039\pm 0.029$&&$1.097\pm 0.006$ &&$1.052\pm 0.013$ && $1.066\pm 0.014$&\cr
&& $\eta_8$&& $0.579$        &&$0.579$         &&$0.557\pm 0.011$ &&$0.572\pm 0.013$ && $0.580\pm 0.012$&\cr
\tablerule
&& $\eta_u$&& $0.777$        &&$0.719$         && $0.764\pm 0.006$&& $0.722\pm 0.010$&& $0.728\pm 0.009$&\cr
&& $\eta_d$&& $-0.333$       &&$-0.321$        &&$-0.320\pm 0.008$&&$-0.320\pm 0.009$&&$-0.325\pm 0.009$&\cr
&& $\eta_s$&& $-0.067$       &&$-0.090$        &&$-0.075\pm 0.008$&&$-0.007\pm 0.007$&&$-0.106\pm 0.008$&\cr
\tablerule
&& $a_0$   &&$0.183\pm 0.030$&&$0.284\pm 0.012$&&$0.183\pm 0.013$ &&$0.255\pm0.006$  &&$0.250\pm0.007$  &\cr
\tablerule
\tablerule
}}}\hfil\medskip
\centerline{\vbox{\hsize= 380pt \raggedright\noindent\footnotefont
Table~1: Best--fit values of the first moments for the data and
pseudodata fits discussed in the text.
}}
\endinsert
We proceed to fit a global set of data which includes the original
neutral--current data as well as the generated charged--current
data. We assign to the generated data the estimated statistical
errors, and fit including statistical errors only. We do not attempt
an estimate of the experimental systematic uncertainties, which are
very difficult to anticipate: our results are meant to provide a
benchmark for the best possible situation of negligible systematics,
and set a target of performance for the planned detectors. The errors
assigned to the neutral--current data are instead obtained as in our
original fits, by adding in quadrature the statistical and systematic
errors given by the various experimental groups: a fit including
statistical errors only to these data would be very hard to achieve,
and somewhat misleading, since the data themselves are affected by the
systematics, which cannot be assumed to be absent. Since the
statistics is dominated by the generated data, the final errors on the
various parameters should be taken as estimates of statistical errors
only.

The fits are performed adopting the same functional form and
parameters as in the original fit for the C--even parton
distributions, except that the normalization of the octet C--even
distribution $\eta_8$ is now also fitted.  For the C--odd parton
distributions, we add six new parameters, namely the normalizations of
the up, down and strange C--odd distributions, and three small--$x$
exponents $\alpha$ (corresponding to an $x^\alpha$ small--$x$
behaviour). The shape is otherwise taken to be the same as that of the
C--even quark distributions.  The exponents $\alpha$ are included
among the refit parameters in order for the refitting procedure to be
more realistic, in view of the fact that the generated data are
distributed around the best fit with appropriate errors, but do not
coincide with it. The charm distribution is assumed to vanish below
threshold, and to be generated dynamically by perturbative evolution
above threshold.

The best--fit values of all the normalization parameters are shown in
the last three columns of Table~1, where the rows labelled $\eta_u$,
$\eta_d$ and $\eta_s$ now give the best--fit values and errors on the
first moments of $\Delta q^-$ for up, down and strange. Comparison of
these values with those of our original fits leads to an assessment of
the impact of charged--current data on our knowledge of the polarized
parton content of the nucleon.

First, we see that the improvement in the determination of
the polarized gluon distribution is minor.
This is due to the fact that the gluon distribution is determined by
scaling violations, and thus a precise determination is only possible
using data which cover a wide range of values of $Q^2$. Lacking this,
the availability of charged--current data {\it per se\/} does not
help. Furthermore, since the gluon decouples from $g_5$, the
corresponding data do not have any effect on the determination of the
gluon distribution. In the `skyrmion' and `instanton' fits, where the
charged--current data have been generated assuming a vanishing
polarized gluon component, the refitted gluon is nevertheless nonzero,
though significantly smaller than in the `anomaly' fit. This is due to
the fact that the currently available (neutral--current) data are also
included in the refit, and underlines the fact that these data support
a gluon distribution which differs significantly from zero. The
charged--current data can only have a limited impact in modifying this
conclusion for the same reason why they lead to a modest improvement
in the precision of the determination of the gluon.

Let us now consider the C--even quark distributions. It is immediately
clear that the precision on the singlet quark first moment is very
significantly improved by the charged--current data: the error on the
first moment of $\Delta\Sigma^+$ is now of a few percent in comparison
to about 10\% with neutral--current DIS. This follows from the fact
that, up to subleading corrections, the singlet quark is directly
measured by the combination eq.~\sig\ of charged--current structure
functions. This improvement is especially significant since the
determination of $\eta_\Sigma$ no longer requires knowledge of the
SU(3) octet component, unlike that from neutral--current DIS, and it
is thus not affected by the corresponding theoretical uncertainty.
With this accuracy, the relatively larger value of the singlet quark
component found in the `anomaly' scenario (in the AB scheme) can be
experimentally distinguished from the smaller value found in other
scenarios, at the level of several standard deviations.  In other
words, thanks to the charged--current data, it is possible to
experimentally refute or confirm the anomaly scenario by testing the
size of the scale--independent singlet quark first moment.
Correspondingly, the improvement  in knowledge of the gluon first
moment,  although modest, is sufficient
to distinguish between the two scenarios.

The determination of the singlet axial charge is improved by an amount
comparable to the improvement in the determination of the singlet
quark first moment. Its vanishing could thus be established at the
level of a few percent. The determination of the isotriplet axial
charge is also significantly improved: the improvement is comparable
to that on the singlet quark, and due to the availability of the
triplet combination of charged--current structure functions
eq.~\bjsr. This would allow an extremely precise test of the Bjorken
sum rule, and accordingly a very precise determination of the strong
coupling.  Finally, the octet C--even component is now also determined
with an uncertainty of a few percent.  Therefore, the strange C--even
component can be determined with an accuracy which is better than
10\%. Comparing this direct determination of the octet axial charge to
the value obtained from baryon decays would allow a test of different
existing models of SU(3) violation~\sutr.

Coming now to the hitherto unknown C--odd quark distributions, we see
that the up and down C--odd components can be determined at the level
of few percent. This accuracy is just sufficient to establish whether
the up and down antiquark distributions, which are constrained by
positivity to be quite small, differ from zero, and whether they are
equal to each other or not.  Furthermore, the strange C--odd component
can be determined at a level of about 10\%, sufficient to test for
intrinsic strangeness, \ie\ whether the C--odd component is closer in
size to zero or to the C--even component.  The `instanton' and
`skyrmion' scenarios can thus also be distinguished at the level of
several standard deviations.

\topinsert
\vbox{
\epsfxsize=10truecm
\centerline{\epsfbox{pwp.ps}}
\vskip.1truecm
\epsfxsize=10truecm
\centerline{\epsfbox{pwm.ps}}
\vbox{\footnotefont\baselineskip6pt\narrower\noindent
Figure 10: The structure functions
$g_1^{W^+}$, $g_5^{W^+}$ (top) and $g_1^{W^-}$, $g_5^{W^-}$ 
(bottom) for a proton target.}}
\medskip
\endinsert
Of course, only experimental errors have been considered so far. In
Ref.~\cracovia\ it has been shown that theoretical uncertainties on
first moments are dominated by the small--$x$ extrapolation and
higher--order corrections. The error due to the small--$x$
extrapolation is a consequence of the limited kinematic coverage at
small $x$. This will only be reduced once beam energies higher
than envisaged in this paper will be achieved; otherwise,  
this uncertainty could
become the dominant one and hamper an accurate
determination of first moments.
 On the other hand, the
error due to higher order corrections could be reduced, since it is
essentially related to the fact that available neutral--current data
must be evolved to a common scale, and also errors are
amplified~\spinrevth\ when extracting the singlet component from
neutral--current data because of the need to take linear combinations
of structure functions. Neither of these procedures is necessary if
charged--current data with the kinematic coverage considered here are
available.

\topinsert
\vbox{
\epsfxsize=10truecm
\centerline{\epsfbox{dwp.ps}}
\vskip.1truecm
\epsfxsize=10truecm
\centerline{\epsfbox{dwm.ps}}
\vbox{\footnotefont\baselineskip6pt\narrower\noindent
Figure 11: The structure functions
$g_1^{W^+}$, $g_5^{W^+}$ (top) and $g_1^{W^-}$, $g_5^{W^-}$ 
(bottom) for a deuterium target.}}
\medskip
\endinsert
The best--fit structure functions corresponding to the `anomaly' refit
(third column of Table~1) are displayed as functions of $x$ at the
scale corresponding to the bin 4~GeV$^2\leq Q^2\leq$~8~GeV$^2$, and
compared with the data in Figs.~10-11 (a few data points with large
error bars, although included in the fits, are not shown in the
figures). Notice that the structure functions $g_1$ and $g_5$ always
have opposite signs because the (dominant) quark component in $g_1$
and $g_5$ has the opposite sign, while the antiquark component has the
same sign. For comparison, we also display the structure functions at
the initial scale of the fits, and at a high scale. The good quality
of the fits is apparent from these plots.

\topinsert
\vbox{
\epsfxsize=10truecm
\centerline{\epsfbox{deltau.ps}}
\vskip.1truecm
\epsfxsize=10truecm
\centerline{\epsfbox{deltaub.ps}}
\vbox{\footnotefont\baselineskip6pt\narrower\noindent
Figure 12: 
The combinations of structure functions of eq.~\flavcoma, and the 
corresponding parton distributions.
}}
\medskip
\endinsert
Given the poor quality of current knowledge of the shape of polarized
parton distributions, it is difficult to envisage detailed scenarios
and perform a quantitative analysis of the various shape parameter, as
we did for first moments. However, it is possible to get a rough
estimate of the impact of charged--current data on our knowledge of
the $x$ dependence of individual parton distributions by considering
the combinations of structure functions given in
eqs.~\flavcoma,\flavcomb, which, at leading order, are directly
related to individual parton distributions. In Figs.~12,13 we show,
respectively, the combinations of eq.~\flavcoma\ and \flavcomb\ for a
proton target, together with the pseudodata for the same combinations
of structure functions, in the bin 4~GeV$^2\leq Q^2\leq$~8~GeV$^2$. In
each figure we also display the two parton distributions which
contribute at leading order to the relevant combination of structure
functions at $Q^2=7$~GeV$^2$, as well as $(\as/\pi)\Delta g$ at the
same scale.

\topinsert
\vbox{
\epsfxsize=10truecm
\centerline{\epsfbox{deltad.ps}}
\vskip.1truecm
\epsfxsize=10truecm
\centerline{\epsfbox{deltadb.ps}}
\vbox{\footnotefont\baselineskip6pt\narrower\noindent
Figure 13: 
The combinations of structure functions of eq.~\flavcomb, and the 
corresponding parton distributions.
}}
\medskip
\endinsert
Let us consider the upper plot in Fig.~12. It is apparent that the
expected statistical accuracy is very good for all data with $x>0.1$.
This suggests that an accurate determination of the shape of $\Delta
u+\Delta c$ is possible. Furthermore, it is also clear that $\Delta c$
(dotted curve) is extremely small compared to $\Delta u$ (solid
curve). However, we observe that the difference between the $\Delta u$
distribution (solid) and the data is of the order of 15\% to 20\% for
all $x$ below 0.4. This difference is entirely due to next-to-leading
corrections. Specifically, the gluon contribution (dot-dashed curve),
which in the AB scheme spoils the leading order identification of
the quark parton distribution with the structure function [see
eq.~\schch], is small but non negligible.  Because the various
contributions to next-to-leading corrections (in particular the gluon
distribution) are affected by sizable theoretical
uncertainties~\cracovia, this implies that $\Delta u$ can only be
determined with an error which is considerably larger than the
experimental one. At larger scales, one expects the subleading
corrections to coefficient functions to be smaller and smaller, while
a residual gluon contribution persists, because of the axial
anomaly~\anom.

A similar analysis of the lower plot of Fig.~12 tells us that a
determination of the shape of $\Delta \bar{u}$ is essentially
impossible.  This combination of structure functions is the preferred
one for a determination of the charm distribution, since perturbatively
we expect $\Delta c=\Delta \bar{c}$, and $\Delta \bar u$ is much
smaller than $\Delta u$. Nevertheless, it is apparent from this figure
that even in this case a determination of the charm distribution
is out of reach.

A study of the down quark and antiquark distributions can be similarly
performed looking at Fig.~13.  The conclusion for $\Delta d$ is
similar, although perhaps slightly less optimistic, to that for
$\Delta u$: a reasonable determination of its shape is possible, but
with sizable theoretical uncertainties.  The lower plot shows that no
significant information on the shape of $\Delta \bar d$ can be
obtained from this analysis.

\newsec{Conclusions and Outlook}
We presented in this paper a self-contained review of the
next-to-leading order formalism for the description of
charged--current polarized DIS, discussing in particular the impact of
next-to-leading order corrections to the evolution equations and to
the relations between polarized structure functions and polarized
parton densities.  We discussed the theoretical constraints on
individual quark and antiquark polarized distributions emerging from
positivity requirements of physical cross-sections. We found that
$\Delta \bar{u}(x)$ and $\Delta \bar{d}(x)$ are constrained to be much
smaller than the valence polarizations $\Delta {u}(x)$ and $\Delta
{d}(x)$, respectively; $\Delta \bar{s}(x)$ and $\Delta {s}(x)$ are
instead allowed to have similar size.

We then focused the phenomenological applications to the case of
neutrino scattering, which has recently become an intriguing
possibility in the context of the neutrino factory facilities being
explored world-wide. After evaluating the statistical uncertainties
which are expected to be achievable in the measurement of polarized
structure functions, we studied the impact of such accuracies on the
theoretical models describing the structure of the proton spin.  In
the case of the C--even distributions, our results indicate that the
singlet, triplet and octet axial charges can be measured with
accuracies which are up to one order of magnitude better than the
current uncertainties. In particular, the improvement in the
determination of the singlet axial charge would allow a definitive
confirmation or refutation of the anomaly scenario compared to the
`instanton' or `skyrmion' scenarios, at least if the theoretical
uncertainty originating from the small--$x$ extrapolation can be kept under
control. The measurement of the octet axial charge with a few percent
uncertainty will allow a determination of the strange contribution to
the proton spin better than 10\%, and allow stringent tests of models
of SU(3) violation when compared to the direct determination from
hyperon decays.

In the case of C--odd distributions, the up and down components can be
determined at the level of few percent, allowing the measurement of an
antiup and antidown polarization at levels small enough to be still
compatible with the positivity constraints.  The strange C--odd
component can be measured at the level of 10\%, sufficient to test for
instrinsic strangeness, and thus distinguish between `skyrmion' and
`instanton' scenarios at a level of several standard deviations.

We also studied the prospects for the determination of the shape of
the polarized distributions.  If one were to assume the leading--order
relation between structure functions and polarized parton densities,
the statistical power of the neutrino factory data would pin down the
shape of $\Delta u(x)$ with a few-percent precision. Most of this
accuracy is however lost when next-to-leading order corrections are
taken into account, which mix the quark and gluon distributions. The
impact of the gluon distribution on the extraction of the $\Delta
u(x)$ shape is at the level of 15-20\%, and any uncertainty on $\Delta
g(x)$ will be reflected on this extraction accordingly.  The results
for $\Delta d(x)$ are similar to those for $\Delta u(x)$, while no
significant shape information can be obtained for the sea
distributions. In the case of $\Delta s(x)$ one may be able to use
semi-inclusive measurements, with tagged charm quarks in the final
state, but this study remains to to be done.

Our analysis did not try to incorporate any estimate of the systematic
experimental uncertainties, nor of the theoretical systematics induced
by the separation of higher-order and higher-twist contributions, and
small-$x$ extrapolation. Likewise, we did not try to include the new
information on the spin structure of the proton and on polarized
parton densities which might become available through future
experiments at CERN, DESY and RHIC. Our study confirms nevertheless
the expectation that polarized DIS experiments at a neutrino factory
will provide invaluable information on the structure of the proton,
provided the experimental systematics will be able to match the
statistical one. At the same time, our work stresses the known fact
that QCD corrections to the parton-level picture are large, and that a
neutrino factory alone may not be able to fully disentangle the shape
of individual quark flavors unless a firmer knowledge of the polarized
gluon is achieved.

\bigskip
{\bf Acknowledgements}: We thank M.~Anselmino, R.~D.~Ball, P.~Gambino,
D.~Harris, J.~Lichtenstadt, K.~McFarland and F.~Zomer for discussions,
and A.~Blondel for providing us with the code describing the neutrino
beam.  This work was supported in part by EU TMR contract
FMRX-CT98-0194 (DG 12 - MIHT).

\vfill\eject
\listrefs
\bye